\newcommand{\psipdc}{\psi_{\rm PDC}}   		
\newcommand{\psisfg}{\psi_{\rm SFG}}   		
\newcommand{\psiout}{\psi_{\rm meas}}           
\newcommand{\UV}{F_{\rm PDC}}  			
\newcommand{\Sicoh}{F_{\rm SFG}^{(\rm coh)}}  			
\newcommand{\Si}{F_{\rm SFG}}                   
\newcommand{\Deltapdc}{\Delta^{\rm pdc}}                   
\newcommand{\Deltasfg}{\Delta^{\rm sfg}}                   
\newcommand{\psia}{\psi_{\rm meas}}   		
\newcommand{\psib}{\psi_{\rm meas}}           
\newcommand{\Icoh}{I_{\rm SFG}^{\rm coh}}           
\newcommand{\vka}{\vec{w}}			
	
\newcommand{\vkap}{\vec{w}\,'}

\newcommand{\vxi}{\vec{\xi}}				
\newcommand{\vxip}{\vec{\xi}\,'}

\newcommand{\im}{i}

\newcommand{\x}{\vec{x}}

\newcommand{\sinc}{{\rm Sinc}}
\newcommand{\q}{\vec{q}}
\newcommand{\qp}{\vec{q}\,'}

\newcommand{\xp}{\vec{x}\,'}

\newcommand{\nn}{\nonumber}
\newcommand{\bsub}{\begin{subequations}}
\newcommand{\esub}{\end{subequations}}
\newcommand{\beq}{\begin{equation}}
\newcommand{\eeq}{\end{equation}}
\newcommand{\beqa}{\begin{eqnarray}}
\newcommand{\eeqa}{\end{eqnarray}}
\newcommand{\beql}{\begin{subequations}\begin{eqnarray}}
\newcommand{\eeql}{\end{eqnarray}\end{subequations}}
\documentclass[pra,twocolumn,aps,nofootinbib,amsmath,showpacs]{revtex4}
\usepackage{amsmath} 
\usepackage{graphicx}
\usepackage{float}   
\usepackage{verbatim}   
\usepackage{braket}
\usepackage{footmisc}
\begin{document}
\title{Disclosing the spatio-temporal structure of PDC entanglement through frequency up-conversion}
\author{E.~Brambilla$^1$, O. Jedrkiewicz$^{1,2}$, L.~A.~Lugiato$^1$,  and A.~Gatti$^{1,2}$}
\affiliation{$^1$ CNISM and Dipartimento di Scienze Fisiche e Matematiche\text{,} Universit\`a dell'Insubria, Via Valleggio 11 Como, Italy, \\
$^2$ CNR, Istituto di Fotonica e Nanotecnologie, Piazza Leonardo da Vinci 4, Milano, Italy}
\begin{abstract}
In this work we propose and analyse a scheme where the full spatio-temporal correlation
of twin photons/beams generated by parametric down-conversion is detected by using its inverse process, i.e.
sum frequency generation. Our main result is that, by imposing independently a temporal delay $\Delta t$ and a transverse
spatial shift $\Delta x$ between two twin components of PDC light, the up-converted light intensity provides information 
on the correlation of the PDC light in the full spatio-temporal domain, and should enable the reconstruction of the peculiar X-shaped
structure of the correlation predicted in \cite{gatti2009,caspani2010,brambilla2010}.
Through both a semi-analytical and a numerical modeling of the proposed optical system, we analyse the feasibility of the experiment and identify the
best conditions to implement it. In particular, the tolerance of the phase-sensitive measurement against the presence of dispersive elements,
imperfect imaging conditions and possible misalignments of the two crystals is evaluated.
\end{abstract}
\maketitle

\section*{Introduction}
\label{sec:intro}
Recent theoretical investigations \cite{gatti2009,caspani2010,brambilla2010}  outlined a peculiar spatio-temporal geometry of the biphoton correlation characterizing the entanglement of twin beams generated by parametric down-conversion (PDC). In collinear phase matching conditions, the biphotonic correlation displays a X-shaped geometry as a function of the relative spatial and temporal coordinates.
This structure is non-factorable in space and time, offering thus the relevant possibility of manipulating the temporal bandwidth of the  entanglement of twin photons by acting on their spatial degrees of freedom. 
The name "X-entanglement" was used \cite{gatti2009} to describe this geometry. 
A  key feature that emerged was the extreme spatial and temporal localization of the biphotonic correlation, on the micrometer and femtosecond range, respectively, which is present only when twin photons are detected in the near field of the PDC source. 
This feature allows in principle the generation of  ultra-broadband temporally entangled photons,  via a proper control of their spatial degrees of freedom. We also showed that these features of  X-entanglement persist in the high gain regime of PDC, where stimulated down-conversion becomes the main source of twin photon pairs.
A detailed study  can be found in  \cite{gatti2009,caspani2010} for type I phase-matching, and in \cite{brambilla2010} for type II phase-matching.

The goal of the present work is a careful theoretical investigation of a scheme based on the use of sum frequency generation (SFG) as a tool to explore the predicted X-shaped geometry of PDC entanglement. 
An experiment based on this scheme is currently under development at the Insubria University in Como \cite{jedr2011,jedr2012}. 

At low gains, a prominent way to probe the twin photon correlation is the  Hong-Ou- Mandel (HOM) detection scheme \cite{HOM}.
However, the experiment in Como works in the high gain regime of PDC, where
the visibility of a HOM dip would be exceedingly low.  
We consider therefore an alternative detection scheme based on the  SFG process 
taking place in a second nonlinear crystal, onto which the PDC source is imaged. 
This second crystal operates
as an ultrafast optical correlator, the up-converted field containing the information about the correlation of the injected source.

Recent experimental works  \cite{dayan2005,dayan2007, peer2005b,odonnel2009, harris2007} used SFG to test 
the twin photon/twin beam correlation in the purely {\em temporal} domain, by imposing a controlled {\em temporal} delay between the twin photons. 
A careful theoretical modeling of such schemes,  and of related schemes based on two-photon absorption \cite{dayan2004},  has been developed in \cite{dayan2007} . However, this analysis is restricted to models considering only the temporal degrees of freedom of light.  A fully spatio-temporal model for PDC/SFG has been considered in \cite{odonnel2009}, however this model is in turn valid only in the coincidence count regime of PDC.  
\par
The aim of the experiment being developed in Como is to use SFG  for exploring the PDC correlation in the whole spatio-temporal domain, by manipulating independently the temporal and the spatial degrees of freedom of the twin beams.
To this end,  we develop here a fully spatio-temporal model in order to describe the non linear processes taking place in the two crystals and the propagation between the crystals. This model is valid both in the high gain of PDC, where the experiment in Como is being performed, and in the low-gain regime, where twin photons can be resolved by coincidence counting. 

We demonstrate that the proposed scheme allows in principle the reconstruction of the  X-shaped spatio-temporal correlation of twin beams/ twin photons, and we identify the best conditions under which an experimental observation can be performed. In particular,  we analyse  important issues such as the visibility of the information thereby obtained  (which becomes crucial in the high gain regime of PDC), and the tolerance of the scheme with respect to common experimental imperfections,  such as errors in the imaging scheme that maps the PDC light onto the SFG crystal or misalignments of the two nonlinear crystals. 

Besides being necessary  to describe the detection of the spatio-temporal X-correlation, we remark that our fully spatio-temporal model  for PDC and SFG  is also crucial to interpret  the results of similar schemes aimed at   exploring the purely temporal correlation of twin photons, at least when the temporal bandwidths in play are large. This statement,  further demonstrated by the experiment reported in \cite{jedr2012}, originates from the intrinsic nonfactorability of spatial and temporal degrees of freedom of the PDC correlation, which implies  that the temporal properties of twin photons depend also on the way their spatial degrees of freedom are manipulated before being up-converted (by the use e.g. of pupils that restrict the acceptance angle). A clear example is given in Sec. \ref{sec:fragility}b, where we show that the effect of spatial diffraction introduced by an imperfect imaging scheme is a temporal broadening of the correlation, in addition to the more obvious spatial broadening. 
 
The paper is organized as follows. 
Sec.\ref{sec:general_scheme} illustrates the proposed  scheme , while in  Sec. \ref{sec:model} we develop 
a modeling of the setup. In  Sec. \ref{sec:SFGfield} we derive a general solution for the coherence function of the generated SFG light, so that in  Sec\ref{sec:retrieve_psi}  we are able to 
demonstrate how the full spatio-temporal  X-correlation can  
be retrieved by monitoring the SFG light intensity as a function of the temporal delay and the spatial transverse shift between the twin beams. Section \ref{sec:visibility} discusses the issue of the visibility of the information, while Sec.\ref{sec:fragility} analyses the tolerance of the scheme with respect to some common experimental imperfections. 

\section{General description of the scheme} 
\label{sec:general_scheme}
The main features of the scheme we propose and of its  theoretical modeling are the following: \par

i) Both the spatial and the temporal degrees of freedom of the optical fields are taken into account. In order to observe X-entanglement, a large temporal bandwidth (hundreds of nanometers) of PDC radiation needs to be collected \cite{gatti2009,caspani2010}, so that special care will be taken to model broadband field propagation. The spatial bandwidth is also rather large 
so that we assume that optical elements  have large acceptance angles. \par
ii) Our description is valid in any gain regime of PDC, so that it models both the generation and detection of twin-photons and of twin-beams. \par
\begin{figure}
\centering
{\scalebox{.6}{\includegraphics*{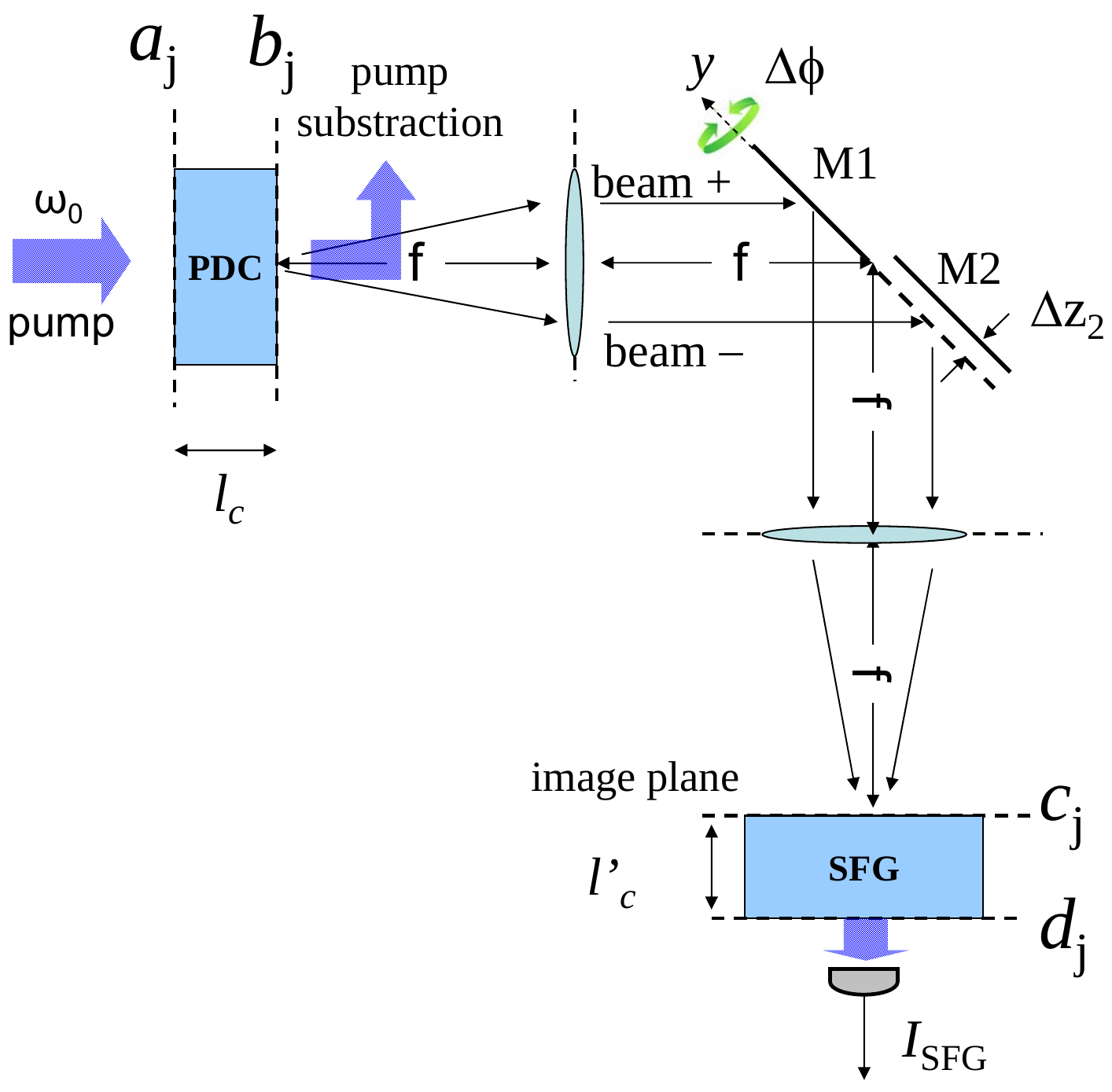}}}
\caption{Proposed scheme for detecting the full spatio-temporal correlation of PDC via sum frequency generation. 
The PDC light generated by a first crystal is imaged onto a second identical crystal, where up-conversion takes place.
A transverse spatial displacement $\Delta x$ and a temporal delay $\Delta t$ are imposed independently on beam $+$ and beam $-$ respectively, by means of rotations/translations of the mirrors M1/M2 placed in the 2-f plane  of the imaging system. The output SFG intensity is monitored as a a function of $\Delta x$ and $\Delta t$ }
\label{fig2}
\end{figure}
iii) A simplified  scheme is illustrated  in Fig.\ref{fig2}.   A first nonlinear crystal, pumped by a broad coherent pump beam, generates PDC radiation. After eliminating the pump beam, the exit face of the PDC crystal is imaged onto the entrance face of a second nonlinear crystal, where the inverse SFG process takes place. This is described in Fig.\ref{fig2} by a 4-f lens imaging system: in a real implementation, however, dispersive optical elements should be avoided as much as possible, since dispersion would drastically deteriorate the phase-sensitive correlation. 
The use of parabolic mirrors in place of lenses is a valid alternative, as we recently demonstrated in \cite{jedr2011,jedr2012}.\par
iv) In order to have an efficient up-conversion process, the SFG  and PDC crystals must be of the same material and  cut for the same  phase-matching conditions, while their lengths can differ. \par
v) We focus on type I PDC, and in the following we shall always refer to the case of  BBO (Beta-Barium-Borate) 
crystal pumped at 527nm in a e-oo phase-matching configuration. The PDC field is then described 
by a single field operator (ordinarily polarized),  while the pump field is extraordinarily polarized. \par
vi) In order to explore the full spatio-temporal structure of PDC entanglement, a key requirement is the ability  to impose  independently  a  temporal delay $\Delta t$ and a transverse  spatial
displacement $\Delta \x$ between the two twin components of PDC radiation.   
The twin components of the signal beams can be manipulated separately in the 2f plane of the imaging device (the far-field plane with respect to both the PDC crystal exit face and the SFG crystal entrance face): the reason is that  for a broad pump beam twin photons are always emitted with opposite transverse wave-vectors $\q$ and $-\q$, so that when a photon is found in the upper half of the 2-f plane,  its twin will always be in the lower half plane.  
We shall denote by beam + and beam -, respectively, the upper and lower portions of the 2f plane, which correspond to the two portions of the PDC radiation having positive $q_y>0$ and negative $q_y<0$ components of the transverse wave vector  along the  $y$-axis in  Fig.\ref{fig2}.\par
vii) Since the PDC correlation  is strongly localized both in space and time  (in the micrometer and femtosecond range respectively \cite{gatti2009,caspani2010,brambilla2010}), the temporal 
and spatial relative displacements of the twin beams  
must be scanned with micrometric precision. 
This can be realized by means of the two plane mirrors M1 and M2 placed in the 2-f plane of the telescopic system. As it will be described in detail in the following , a rotation of mirror M1 by an angle $\Delta \phi$ generates at the SFG crystal input face  a transverse displacement of  beam + of an amount $\Delta x=2 f \Delta \phi $, while a translation of mirror M2 by a distance $\Delta z_2$ generates a temporal delay  $\Delta t=\Delta z_2/c$.
\par
The intensity of the up-converted field generated by the SFG process in the second crystal is then monitored as a function of the relative temporal delay and spatial shift of the twin beams.  The main idea that we will demonstrate is that this quantity is able to give a precise information about the structure of the correlation of twin beams generated in the first crystal and can  be used in  order to reconstruct the shape of this correlation in space and in time.

In the treatment that follows we develop a model that takes fully into account the propagation effects and the phase-matching mechanism that selects the spatio-temporal frequencies in both crystals.

\section{Modeling the optical system} 
\label{sec:model}
In order to describe the scheme of Fig.\ref{fig2}, we consider 
separately the propagation in the PDC crystal (Step I), the linear propagation between the two crystals (Step II) and the up-conversion process in the second crystal (Step III). 
As indicated in Fig.\ref{fig2}, the field operators in the different planes of interest will be labeled with \\
- $a_j$ at the input plane of the PDC crystal; \\  
- $b_j$ at the output plane of the PDC crystal; \\
- $c_j$ at the input plane of the SFG crystal; \\
- $d_j$ at the output plane of the SFG crystal. \\
The $j=0$ index refers to beams with central frequency $\omega_0$ (either the pump field in the first crystal or the up-converted field in the second crystal), the $j=1$ index to the beams of central frequency $\omega_1=\omega_0/2$ (either the down-converted field in the first crystal or the fundamental field in the SFG crystal). 
\par
The description of field evolution along $z$,  the mean propagation axis of the system, will   be performed either in the direct spatio temporal space $(\x, t)$, where $\x \equiv (x,y) $ is the 2D transverse coordinate and $t$ is time, or in the Fourier spatio-temporal domain $(\q, \Omega)$ where $\q $ is the transverse component of the wave-vector and $\Omega$ denotes the frequency offset from the central frequency. \\
For convenience, we shall use a compact notation for these space-time coordinates 
by making the substitutions
\bsub
\label{compact}
\beqa
&&(\x,t) \rightarrow\vxi\\
&&(\q,\Omega) \rightarrow \vka \\
&&\q\cdot\x-i\Omega t \rightarrow\vka\cdot \vxi
\eeqa
\esub

Our analysis will be carried out at two levels\\
1) In the limit where the pump beam driving the PDC process is broad and long enough, we shall adopt the plane-wave pump approximation (PWPA), which allows us to derive analytical or semi-analytical results. This model will be presented in the next sections \ref{sec:PDC}-\ref{sec:visibility} .\\
2)In order to obtain results for a finite pump,  we also developed a full 3D+1 numerical model, based on stochastic simulation of field evolutions. This model will be introduced in Sec. \ref{sec:visibility} 
\subsection*{Step I: propagation in the PDC crystal}
\label{sec:PDC} 

In this section we describe the model we use to describe PDC (also derived in \cite{gatti2003, gatti2009, caspani2010}), and we recall the main features of the space-time correlation of PDC light that we called X-entanglement \cite{gatti2009, caspani2010, brambilla2010}.
\par
The pump and down-converted fields are described by two field operators $b_0$ and $b_1$,  centered around the frequencies $\omega_0$ and $\omega_1= \omega_0/2$, respectively.  Normalization is such that $\langle b_j^{\dagger}(\xi)b_j(\xi)\rangle$ gives the photon number  per unit area and unit time. 
The generation of the PDC field along the nonlinear crystal takes its simplest form in the Fourier domain
\beq
b_j(\vka,z) = \int \frac{  d^3 \xi } {  (2\pi)^{\frac{3}{2} }}  b_j (\xi,z) e^{-i \xi \cdot \vka}  \quad j=0,1 \, . 
\eeq 
where we recall that  $\vka = \q,\Omega$  is the set of 3D Fourier coordinates,   $\xi = (\x, t) $,  while $z$ is the longitudinal coordinate along the mean propagation direction in the crystal. 
Next,  we introduce the slowly varying amplitudes
\bsub
\label{slow_ampl}
\beqa
\bar{b}_1(\vka,z)&=&e^{-\im k_{1z}(\vka)z} b_1(\vka,z)\, , \\
\bar{b}_0(\vka,z)&=&e^{-\im k_{0z}(\vka)z} b_0(\vka,z)\, ,  
\eeqa
\esub
where $k_{jz}(\vka)=\sqrt{k_j(\vka)^2-q^2}$ is the z-component of the wave-vector for the $j$-field.      
These amplitudes vary slowly along the $z$-coordinate, because their evolution is only  due to the nonlinear interaction, since we have subtracted the effect of the fast linear propagation  contained in the phase factor $\exp{[\im k_{jz}(\vka)z]}$. 
We can assume  the pump beam is undepleted by the PDC, so that its evolution is only linear  $\bar{b}_0(\vka,z) = \bar{b}_0(\vka,0)  $ . In the same approximation, the pump operator can be substituted by its c-number amplitude $\bar{b}_0(\vka,0)  \to \alpha_p (\vka)$.
The propagation equation for the signal field contains only first-order z-derivatives and takes the form \cite{gatti2003,gatti2009}
\beqa
\label{propPDC}
\frac{\partial} {\partial z}     \bar{b}_1 (\vka,z )    &=& 
			\sigma\int  \frac{d\vkap}{(2\pi)^{3/2}} 
					\alpha_p (\vka+\vkap) \bar{b}_1^\dagger(\vkap, z ) 
\nn\\  & & \times e^{-i\Delta(\vka, \vkap)z}  \, . 
\eeqa 
Here the phase matching function 
\beq
\label{DeltaPDC}
\Delta(\vka, \vkap)=
k_{1z}(\vka)+k_{1z}(\vkap)
-k_{0z}(\vkap+\vkap)
\eeq
describes  the phase-mismatch between the two generated  signal modes $(\q,\Omega)$, $(\qp,\Omega')$  
and the pump mode $(\q+\qp,\Omega+\Omega')$. Efficient down-conversion takes place only in those modes for which the phase-mismatch is small. 
The coupling constant $\sigma$ is defined by
\beq
\sigma = d_{eff} \sqrt{\frac{\hbar\pi^3\omega_0^3}{4\epsilon_0 n_0 n_1^2c^3}}  
\eeq
where $d_{eff}$ is the effective second order susceptibility of the nonlinear crystal, $n_0$ and $n_1$
are the refraction indexes  at the central frequencies  $\omega_0$ and $\omega_1 $. 
\par

Eq.\eqref{propPDC} can be analytically solved in the plane-wave pump approximation, 
$\alpha_p(\vka+\vkap)\rightarrow(2\pi)^{3/2}\bar{\alpha}_p\delta(\vka+\vkap)$, where $\bar{\alpha}_p$ denotes the pump field
peak value in direct space. As analyzed in detail in \cite{caspani2010}, such an approximation holds as long as the pump beam waist and duration are 
larger than the spatial transverse displacement and temporal delay experienced by the pump and signal beams along the crystal because of walk-off
and group velocity dispersion (in the example of a 4mm long BBO, a pump pulse with a waist larger than $\sim 250\,\mu$m and a duration above $\sim 200\,$fs satisfies this condition).
The solution is expressed by a unitary transformation linking  
the field
operators at the output face of the crystal, $b_1(  \vka)\equiv b_1(\vka,z=l_c)$, to those at at the input face, $a_1(\vka)\equiv b_1(\vka,z=0)$:
\beq
b_1( \vka )=U(\vka)a_1(\vka)+V(\vka)a_1^\dagger(-\vka)\;,
\label{inputoutput1}
\eeq	 
The explicit expression of the  functions $U$ and $V$ can be found e.g. in \cite{caspani2010}. Here we notice that they depend on $\vka$ only through the plane-wave phase mismatch 
\beq
\label{deltaPW}
\Deltapdc  (\vka)\equiv \Delta (\vka, -\vka) = k_{1z}(\vka)+k_{1z}(-\vka) -k_{0}
\eeq
where  $k_0= n_0 \omega_0/c$ is the wave number of the pump. 
All the properties 
of the PDC light are described by the following second-order field correlation functions
\bsub
\label{corr_PDC}
\beqa
&&\langle b_1^{\dagger}(\vka)b_1(\vkap)\rangle=\delta(\vka-\vkap)|V(\vka)|^2\label{corr_PDCa}\\
&&\langle b_1(\vka)b_1(\vkap)\rangle=\delta(\vka+\vkap)  U(\vka)V(-\vka)\label{corr_PDCb}\;,
\eeqa
\esub
In particular, from Eq. \eqref{corr_PDCb}, we see that the function 
\beq 
\UV(\vka) \equiv  U(\vka)V(-\vka)\
\eeq
represents the probability amplitude that  a pump photon at  $(\q=0,\Omega=0)$ 
is down-converted into a pair of phase-conjugated photons  $(\q,\Omega)$ and $(-\q,-\Omega)$.
Its explicit expression is 
\bsub
\label{UV} 
\beqa
\UV(\vka) &=& g e^{ik_0 l_c}\frac{\sinh\Gamma(\vka)l_c}{\Gamma(\vka)l_c}
\left\{\phantom{\frac{a}{b}}\hspace{-.3cm} \cosh\Gamma(\vka)l_c \right. \nn\\
& + & \left.  i\frac{\Deltapdc(\vka)l_c}{2\Gamma(\vka)l_c}
     \sinh\Gamma(\vka)l_c \right\}\;,\\
\Gamma(\vka)l_c
      &=&\sqrt{g^2-\frac{[\Deltapdc(\vka)l_c]^2}{4}}\;,
      \label{Gammalc}
\eeqa
\esub
where $g$ is the dimensionless gain parameter proportional to the pump peak amplitude
\beq
g= \sigma l_c \bar{\alpha}_p. \
\label{g}
\eeq
The other relevant function is the PDC spatio-temporal spectrum
\beq
|V(\vka)|^2= g^2\frac{\sinh^2 [\Gamma(\vka)l_c] }{\Gamma^2(\vka)l_c^2}\
\label{V} 
\eeq
which gives the photon number distribution in the spatio-temporal Fourier domain.\\
We remark  that equations \eqref{propPDC}-\eqref{V} are valid in any gain regime of PDC. In the low gain regime  $g \ll 1$  they describe the down-conversion of pump photons into pairs of signal-idler photons that can be resolved individually.  In the high gain regime $g \sim 1$ stimulated down-conversion becomes important,  and these equations  describe the generation of macroscopic twin beams of light made of "bunched" pairs of twin photons. \\
In particular, in the low-gain regime Eq. \eqref{UV} takes the well known {\sinc} dependence on the phase mismatch:
\beq
\label{UVlow}
\lim_{g \ll 1} \UV (\vka) = g e^{ik_0 l_c} \sinc \frac{\Deltapdc (\vka) l_c} {2} e^ { i \frac{\Deltapdc (\vka)  l_c} {2} } \, .
\eeq 

\par
\begin{figure}
\centering
{\scalebox{.52}{\includegraphics*{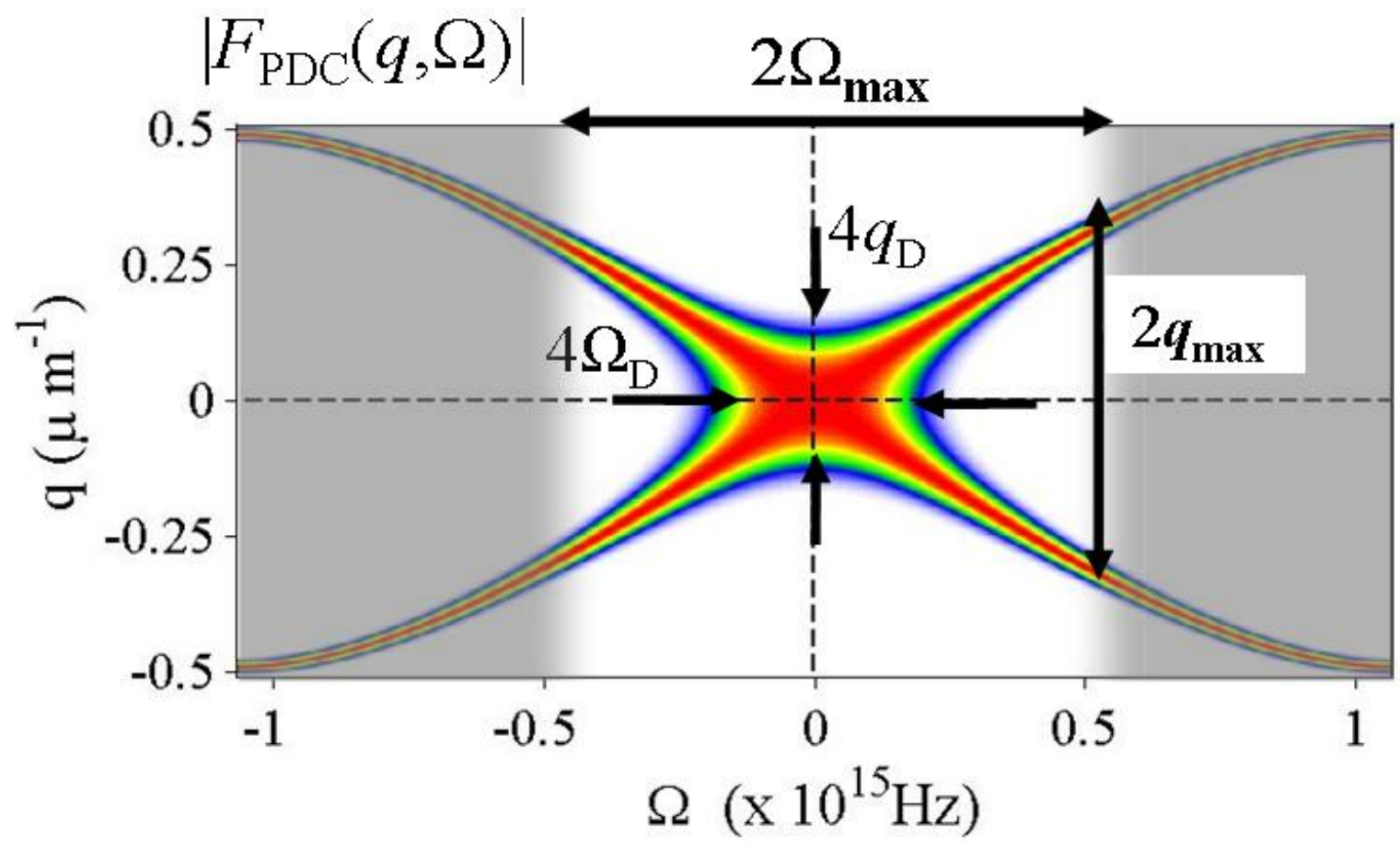}}}
\caption{Plot of $|\UV|$ in the ($q_x, \Omega$)-plane,  for a 4mm long type I BBO pumped at 527nm, cut for collinear phase matching. The unshaded region corresponds to the frequency filter used in the simulation with $2 \Omega_{\rm max}=0.965\times10^{15}$Hz FWHM. 
The bandwidths $\sim 4q_D$ and $\sim 4\Omega$ of the phase-matching of $|\UV|$ along the spatial and the temporal frequency axis are indicated. The parametric gain is $g=8$,  $\Delta_0^{\rm PDC}=0$. 
}
\label{figPM}
\end{figure}

In Ref. \cite{gatti2009} the name X-entanglement was used to describe the shape of the spatio-temporal correlation of the  biphoton amplitude at the crystal output face. The quantity of interest is therefore: 
\beq
\psipdc(\vxi ) = \langle b_1(\vxi) \, b_1(\vxi + \Delta\vxi) \rangle
\label{psi}
\eeq
which in the  stationary plane-wave pump regime depends only on the relative spatial and temporal coordinates 
$\Delta \xi \equiv (\Delta\x, \Delta t)$. It can be expressed \cite{caspani2010} 
as the Fourier transform of the spectral probability amplitude  $\UV(\vka)$ of generating photons in phase conjugate  modes 
$\vka$ and $-\vka$ , that is
\beq
\psipdc(\xi)=
\int  \frac {d\vka }{(2\pi)^{3}} 
e^{i\vxi \cdot\vka}
\UV(\vka)\;,
\label{psi_PDC}
\eeq
Fig. \ref{figPM} shows the behaviour of $|\UV|$ for collinear phase matching, calculated by using  the Sellmaier relations for the refractive indexes \cite{boeuf2000}. As can be inferred from Eqs. (\ref{UV}) and (\ref{UVlow}),  in any gain regime  $\UV (\vka)$  is strongly peaked around the phase matching curves defined by $\Deltapdc(\vka) =0$. The hyperbolic geometry  displayed in the neighborhood of degeneracy can be understood 
by making a  quadratic expansion of  $k_z $ around $(\q=0,\Omega=0)$ 
\beqa
\label{kz}
k_{1z}(\q,\Omega)
\approx k_1+k_1'\Omega+\frac{1}{2}k_1''\Omega^2-\frac{q^2}{2k_1} 
\eeqa
where  $k_1 =k_1(\vka=0)$, $k_1' = d k_1/ d\Omega |_{\vka=0}$ and  $k_1'' = d^{2} k_j/ d\Omega^{2} |_{\vka=0}$ . 
\footnote{A detailed discussion on the range of validity of the approximations (\ref{kz}) 
can be found in \cite{brambilla2010} in the context of  type II phase-matching.} The phase-matching function (\ref{deltaPW}) takes then
the quadratic form
\beq
\label{pm2}
\Deltapdc(\vka)l_c\approx \Delta_0^{\rm PDC} l_c+\frac{q^2}{q_D^2}-\frac{\Omega^2}{\Omega_D^2}
\eeq
where $\Delta_0^{\rm PDC}= 2k_1-k_0$ is the collinear phase-mismatch parameter, and  
\beq
\Omega_D=\sqrt{1/k_1''l_c},\;\;\;\;
q_D=\sqrt{k_1/l_c}
\label{PDC_band}
\eeq
$q_D$ and $\Omega_D$ determine the characteristic scale of variations of $\UV$ along the $\Omega$-axis, at fixed $q$, and along the  $q$-axis at fixed $\Omega$, respectively. They scale with the inverse square root of the crystal length and are generally much smaller than the range of frequencies of the whole PDC emission spectrum (the latter can in principle extend up to the pump optical frequency).
When tuning the crystal for collinear phase matching, i.e for  $\Delta_0^{\rm PDC}=0$,  
phase-matching occurs along  the lines $q/q_D=\pm\Omega/\Omega_D$. 
Under this condition, we notice that the first zeros of the function $\Gamma(\q,\Omega)l_c$ (see Eq.\eqref{Gammalc}) 
along the $\Omega-$ and $q-$axis, evaluated using approximation \eqref{pm2}
\beq
\bar{\Omega}_D=\sqrt{2}(\pi^2+g^2)^{1/4}\Omega_D,\;\;\;
\bar{q}_D=\sqrt{2}(\pi^2+g^2)^{1/4}q_D,
\label{PDC_band_g}
\eeq
provide good estimates of the widths of $|F_{\rm PDC}|$ along those axis, which takes into
account the gain broadening effect in the spectral domain.
For the example shown in Fig.\ref{figPM}, with $g=8$, we have $\bar{\Omega}_D\approx 4 \Omega_D$, $\bar{q}_D\approx 4 q_D$, 
as indicated by the arrows.
\par
\begin{figure}
\centering
{\scalebox{.5}{\includegraphics*{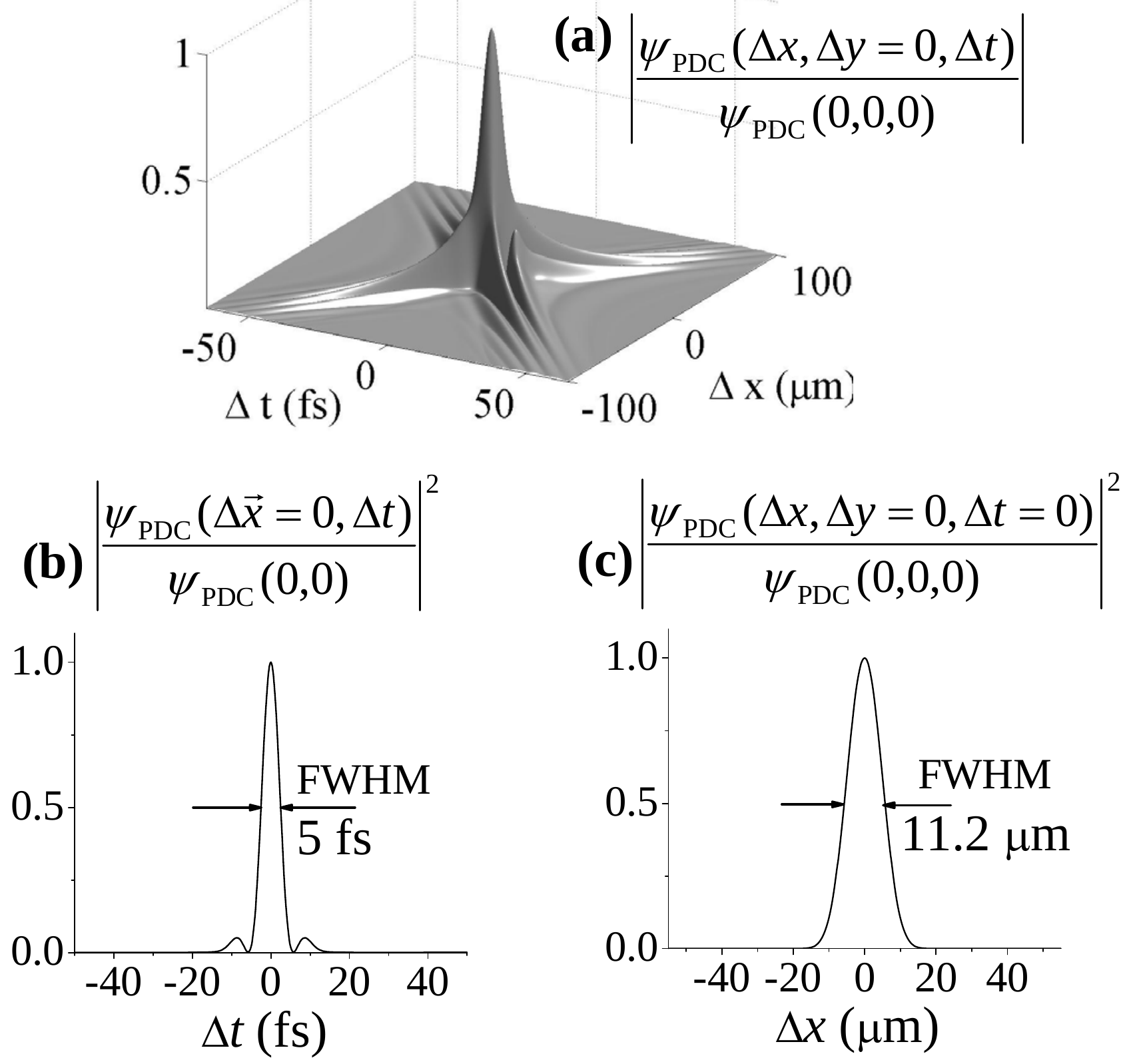}}}
\caption{Example of X-entanglement for the same crystal as in Fig. \ref{figPM}. (a) Plot of the modulus of the biphoton amplitude  as a function of the relative spatio-temporal coordinates $\Delta x, \Delta t$.  Temporal (b) and spatial (c) profile of the central peak of the correlation. Same
parameters as in Fig.\ref{figPM}. 
}
\label{fig1}
\end{figure}

The hyperbolic geometry of  phase matching in the spectral domain turns into a characteristic X-shaped geometry 
for the spatio-temporal biphoton correlation  $\psipdc (\Delta \x, \Delta t)$ (\ref{psi_PDC}). 
Figure \ref{fig1}a  shows the profile of $\psipdc$ as a function of the relative spatial and temporal coordinates. (only one transverse dimension is shown).  
The tails of the structure are oriented along the lines $\Omega_D \Delta t= \pm q_D \Delta x$, a feature expressing a linear relation 
between the temporal delay and the spatial transverse separation acquired by the twin photons when arriving at the crystal output face \cite{gatti2009,caspani2010}, i.e. making use of 
the equations  (\ref{PDC_band}) 
\beq
\label{xt_relation}
\Delta t= \pm  \sqrt{k_1'' k_1} \Delta x
\eeq
As described in detail in \cite{caspani2010}, when tilting the crystal away from collinear phase matching, this linear relation  becomes less and less stringent: as a consequence  the tails of the structure become progressively less visible. 
The $\Delta_0^{\rm PDC}=0$ phase-matching condition represents therefore the optimal 
configuration to observe  the X-entanglement. 
\par
A relevant feature of the X-entanglement is the strong  localization of the central peak of the correlation function, whose cross-sections along the temporal and the spatial axes are shown in the lower part of Fig.\ref{fig1} . 
The width of the peak is in principle determined by the inverse of the full PDC emission bandwidth, or in practice by the bandwidth intercepted in the measurement.  In the example of the figure, we simulated the presence  of a filter in the temporal frequency  
of width   $2\Delta\Omega_{\rm max}=0.965\times 10^{15}$Hz
(indicated by the unshaded region in Fig.\ref{figPM}),  which  
corresponds to a wavelength interval ranging from $830$nm up to the conjugate wavelength $1444$nm. 
The temporal profile of the correlation in Fig.\ref{fig1} has a full width half-maximum 
$\Delta t_{\rm fwhm}\approx 5$fs, close to that of the Fourier transform of a box function of 
width $2\Omega_{\rm max}$. 
The spatial width of the correlation peak, in turn, is determined by the range of spatial frequencies involved in the PDC emission, and typically is on the order of a few microns.

\subsection*{Step II: propagation between the two crystals}
\label{sec:propbetween} 
We assume ideally that the optical setup illustrated in Fig.\ref{fig2} behaves as a perfect imaging system, free from dispersion and losses. 
In the absence of any temporal delay and spatial shift ($\Delta t=0,\Delta x=0$), 
the PDC field at the PDC crystal exit face is then mapped into the SFG crystal entrance face   $c_1(\x,t)=b_1(\x,t)$  \footnote{For simplicity we neglect here and in the following the  spatial reflection with respect to the
$z$-axis $\x\to -\x$, as well as constant phase factors} .  
In order to model the effect of a small rotation of mirror $M_1$ and a small translation of mirror M2  
we assume that the propagation angles in play are such that the paraxial approximation holds. 
\par
Let us first consider the temporal delay applied to the $q_y<0$ Fourier modes of the PDC light (beam -).  By translating the mirror M2 by $\Delta z_2$ the separation between the two lenses in Fig.\,\ref{fig2}  becomes $2f+\Delta z_2$.  The fields at frequency $\omega_1+\Omega$ at the left and right focal planes of the imaging device are related through the algebraic transformation  
(the $\x\rightarrow-\x$ reflection is omitted)
\beq
\label{ff_delay}
c_1(\x,\Omega)=
e^{i\frac{\omega_1+\Omega}{c}\left(1-\frac{|\x|^2}{f^2}\right)
\Delta z_2}
b_1(\x,\Omega)
\eeq
The effect of diffraction due to the additional propagation  $\Delta z_2$ is described, within the paraxial approximation, by 
the quadratic phase factor proportional to $|\x|^2/f^2$. Its effect can be neglected as long as 
\beq
\label{2f_diffraction}
\Delta z\ll\frac{\lambda}{\pi}
\frac{f^2}{|\x|^2}.
\eeq
Referring to the conditions of the experiment \cite{jedr2011,jedr2012}, for
$f\sim 20$cm, $\lambda\sim 1\mu$m and $|\x|\sim w_p \le 1$mm the condition reads $\Delta z\ll 1.2$cm.
In order to explore the X-shaped PDC correlation, delays of a few hundreds femtoseconds at most are sufficient (see e.g. the plot in Fig. \ref{fig1}a), so that in practice condition (\ref{2f_diffraction}) is always fulfilled.
Each Fourier mode of beam - undergoes therefore, within a very good approximation, the 
following diffractionless transformation
\beq
\label{M2_trasl}
c_1(\q,\Omega)=e^{i\frac{\Omega}{c}\Delta z_2}b_1(\q,\Omega)  \;\;\textrm{for}\;\;q_y<0
\eeq
where an inessential constant phase factor has been omitted. 
\par
We now consider the manipulation of beam + through the rotating mirror M1. As shown in App.\ref{appendixA},
a rotation of M1 by a small angle $\Delta \phi$ around a given axis generates a transverse shift $\Delta s=2f\Delta\phi$ of the beam at the entrance face of the SFG crystal (the imaging plane), in the direction orthogonal to the rotation axis. According to Eq. (\ref{M1_xshift}),  
the complete transformation undergone by beam + between the input and output planes of the 4-$f$ telescopic system  
can be written in the Fourier domain in the form (the $\q\rightarrow-\q$ reflection and the minus sign are here omitted
for simplicity)
\beq
\label{M1_rot}
c_1(\q,\Omega)=e^{i\q\cdot\Delta\vec{s}}b_1(\q,\Omega)\;\;\text{for}\; q_y>0
\eeq
where $\Delta \vec{s}$ denotes the transverse shift generated by a rotation of mirror M1
around a generic axis.
Putting together relation (\ref{M2_trasl}) and (\ref{M1_rot}), the 
overall transformation describing propagation from the PDC crystal output face to the SFG crystal input face 
can therefore be synthesized into the following unitary input-output relation
\beq
c_1(\q,\Omega) =
\left[ \Theta(q_y)H_+(\q,\Omega) +\Theta(-q_y)H_-(\q,\Omega)\right] b_1(\q,\Omega) 
\label{inout1}
\eeq
where $\Theta(q_y)$ denotes the step function, equal to 1 for $q_y\ge 0$, to zero for $q_y<0$, and
\beq
\label{transferH}
H_+(\q,\Omega)=e^{i\q\cdot\Delta \vec{s}},\;\;\;H_-(\q,\Omega)=e^{i\frac{\Omega}{c}\Delta z_2}
\eeq
are the transfer functions associated to beam $+$ and beam $-$, respectively.  The identity $\Theta(q_y)+\Theta(-q_y)\equiv 1$
guarantees that the commutation rules are preserved by the transformation.

\begin{figure}[ht]
\centering
{\scalebox{.6}{\includegraphics*{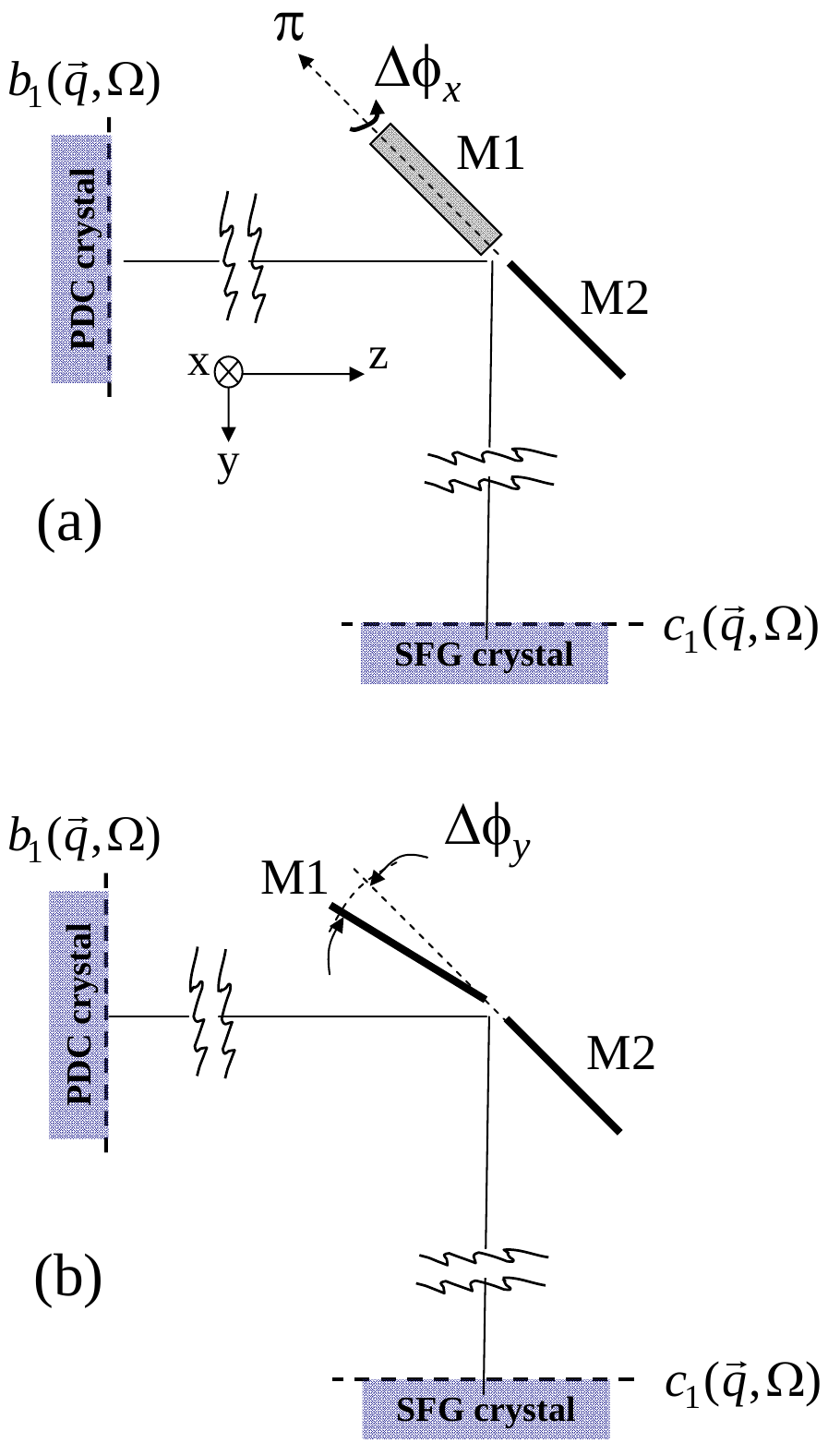}}}
\caption{In configuration (a) the rotation by a small angle $\Delta \phi_x$ of mirror M1 around the $\pi$-axis produces the displacement $\Delta\vec{s}=(\Delta x=2f\Delta\phi_x,0)$ along the $x$-direction at the entrance face of the SFG crystal (orthogonal to the figure plane). In case (b) the rotation is performed around the $x$-axis and produces the displacement  
$\Delta\vec{s}=(0,\Delta y=2f\Delta\phi_y)$ along the $y$-direction.}
\label{fig_rotM1}
\end{figure}
It is important to notice that the correlation measurement depends on the particular choice of the rotation axis of M1, as it will be shown in
Sec.\ref{sec:retrieve_psi}. 
We shall consider explicitly the two configurations illustrated in Fig.\ref{fig_rotM1}: in case a) M1 is rotated around the $\pi$-axis orthogonal 
to the gap between the two mirrors, while in case b) the rotation axis $\pi$ of M1 is coincident with the $x$-axis (orthogonal 
to the figure plane). We have thus 
\bsub
\beqa
&&\Delta \vec{s}=(\Delta x,0),\;\Delta x=2f \Delta\phi_x\;\text{in case (a)}\label{M1a}\\
&&\Delta \vec{s}=(0,\Delta y),\;\Delta y=2f \Delta\phi_y\;\text{in case (b)}\label{M1b}
\eeqa
\label{M1}
\esub
where $\Delta \phi_x$ and $\Delta \phi_y$ denote a small rotation angle applied to the mirror 
around the $\pi$-axis in configuration (a), around $x$-axis in configuration (b).

\subsection*{Step III: propagation in the SFG crystal}
\label{sec:SFG} 
We now model the generation of the up-converted field inside the SFG crystal.  
This process is just the reverse of the down-conversion described in Sec. \ref{sec:PDC}.  Following the same procedure outlined there, we work in the Fourier domain and we introduce the slowly varying amplitudes of the fundamental $d_1$  (carrier frequency $\omega_1$), and second harmonic $d_0$ (carrier frequency $\omega_0$) fields along the SFG crystal
\bsub
\label{slow_ampl2}
\beqa
\bar{d}_0(\vka,z)&=&e^{-\im k_{0z}(\vka)z} d_0(\vka,z)\, \\ 
\bar{d}_1(\vka,z)&=&e^{-\im k_{1z}(\vka)z} d_1(\vka,z)\, , 
\eeqa
\esub
which vary along $z$ only because of the nonlinear interaction, since we have subtracted the effect of linear propagation. Their evolution is described by the following pair of coupled equations:  
\bsub
\label{propSHG}
\beqa
\label{waveSHG2}
\frac{\partial\bar{d_0}(\vka,z)}{\partial z} &=&
     -\sigma   \int\frac{d\vkap}{(2\pi)^{3/2}}  
         \bar{d}_1(\vkap,z ) \bar{d}_1 (\vka-\vkap,z)\nn\\ 
		&&\;\;\;\;\;\;\;\;\;\;\;\;\;\;\;\;\;\;\times e^{i\Delta  (\vkap, \vka-\vkap)z} \, , \\
\label{wavePDC}
\frac{\partial \bar{d}_1 (\vka,z )}{\partial z}&=& 
			\sigma\int\frac{d\vkap}{(2\pi)^{3/2}} 
					\bar{d}_0 (\vka +\vkap ,z) \bar{d}_1^\dagger(\vkap, z ) \nn \\ 
	 &&\;\;\;\;\;\;\;\;\;\;\;\;\;\;\;\;\;\;\times e^{-i\Delta(\vka, \vkap)z}  \, . 
\eeqa 
\esub
Equation \eqref{waveSHG2} describes all the up-conversion processes where a pair of fundamental photons in modes $\vkap$,  $\vka-\vkap$ are up-converted into a SFG photon in mode $\vka$. Accordingly, in this equation $\Delta (\vkap, \vka-\vkap)=
k_{1z}(\vkap)+k_{1z}(\vka-\vkap)
-k_{0z}(\vka)$ represents the phase mismatch of such a process in the second SFG crystal. 
Equation\eqref{wavePDC} is obviously just the same as Eq. \eqref{propPDC} describing the down-conversion process. 
\par
Equations (\ref{propSHG}) need to be considered together with the initial conditions at the crystal input face. For the fundamental field we have
\beq
\bar{d_1} (\vka,z=0) = c_1 (\vka) 
\eeq
where $c_1$ is the field down-converted in the first PDC crystal, after Step I and II. For the second harmonic field, we 
assume that the pump field is completely eliminated after the PDC crystal, so that it is in the vacuum state at the SFG crystal input.

We now assume that the SFG crystal is short enough so that only a small fraction of the PDC light is up-converted.
In these conditions,  the fundamental field remains basically unchanged during propagation, and 
Eq.(\ref{propSHG}) can be solved with a perturbative approach similar to that 
used in the low gain regime of PDC in Ref.\cite{caspani2010} (see App.A therein).  In this way,   
we obtain an explicit expression that links the operators at the SFG crystal output plane to those at the SFG input plane:
\begin{widetext}
\bsub
\beqa
d_1(\vka) &\equiv&   d_1(\vka, l_c')
=   e^{\im k_{1z}(\vka) l_c' }    c_1(\vka) \, , \\
d_0(\vka) &\equiv&   d_0(\vka, l_c')
= e^{\im k_{0z}(\vka) l_c'}   \left[ 
c_0(\vka) 
-
\int\frac{d\vkap}{(2\pi)^{3/2}}
         c_1(\vka-\vkap)
         c_1(\vkap)
		\Si (\vka-\vkap,\vkap)        \right] \, , 
\label{inputoutput_sfg}
\eeqa
\esub
\end{widetext}
where
\beq
\label{Sidef}
 \Si (\vka,\vkap) =\sigma l_c'
e^{i\frac{\Delta(\vka, \vkap)l_c'}{2}} 
                        \sinc\frac{\Delta(\vka, \vkap)l_c'}{2} \, . 
\eeq
The function $\Si(\vka,\vkap)$ can be interpreted as the probability amplitude density
\footnote{Its square modulus $|\Si(\vka,\vkap)|^2$ gives the probability per unit of spectral bandwidth.}
that a pair of photons in the fundamental modes $\vka\equiv(\q,\Omega)$
and $\vka'\equiv(\q',\Omega')$ are up-converted into the second-harmonic mode $\vka+\vka'\equiv(\q+\q',\Omega+\Omega')$:
this up-conversion probability is non negligible only for those pair of modes for which the phase-mismatch
$\Delta(\vka, \vkap)l_c'\approx 0$. As it can be expected,
the probability amplitude for such a process is formally identical to that of the reverse process of down-conversion. 
Eq.\eqref{Sidef} has indeed the same form as Eq. \eqref{UVlow}, the main difference being
that Eq. \eqref{UVlow} describes PDC only in the plane wave pump limit,
where the only allowed down-conversion processes are those leading to twin
photons in modes $\vka$ and $\vkap=-\vka$.  
\section{General solution: the SFG coherence function}
\label{sec:SFGfield}
We now put together the chain of field transformations presented in the previous section.  Our main goal will be to evaluate the intensity of the SFG field, but we start from a more general result, i.e.  the coherence function of the SFG field in the spectral domain,  evaluated at the output
face of the SFG crystal
\beq 
\label{CSFGdef}  
{\cal C}_{\rm SFG}( \vka_0,\vkap_0)
=
\langle
d_0^{\dagger}(\vka_0)d_0(\vkap_0)
\rangle \, .
\eeq
Inserting the input output relation (\ref{inputoutput_sfg}) for the SFG crystal inside this expression, we obtain an equation that links the SFG coherence function to the correlation functions of the fundamental field $c_1$ at crystal input face. 
These correlations can be calculated by 
using relations (\ref{corr_PDC}) together with the identities $|H_+(\vka)|^2=|H_-(\vka)|^2\equiv 1$ and
$\Theta(q_y)+\Theta(-q_y)\equiv 1$.  We obtain the following expression 
\bsub
\beqa	
\langle c_1^{\dagger}  (\vka)c_1 (\vkap)\rangle
&=&\delta(\vka-\vkap)|V(\vka)|^2\\	
\langle c_1(\vka)c_1(\vkap)\rangle
&=& \delta(\vka+\vkap)U(\vka)V(-\vka) \nn \\
&\times & \left[ \Theta(q_y)H_+(\vka)H_-(-\vka) \right. \nn \\
&+& \left. \Theta(-q_y)H_+(-\vka)H_-(\vka)\right] 	\label{c1c1}
\eeqa
\esub
Using these relations 
we obtain after some manipulations the following result
\beq
\label{sfg_correl}
{\cal C}_{\rm SFG}(\vka_0,\vkap_0)
={\cal C}_{\rm SFG}^{\rm (inc)}(\vka_0,\vkap_0)+{\cal C}_{\rm SFG}^{\rm (coh)}(\vka_0,\vkap_0) 
\eeq
with
\begin{widetext}
\bsub
\label{sfg_correl2}
\beqa
&&{\cal C}_{\rm SFG}^{\rm (inc)}(\vka_0,\vkap_0)=
2\delta(\vka_0-\vkap_0)
 \int\frac{d\vka}{(2\pi)^{3}}
 |V(\vka_0-\vka)|^2|V(\vka)|^2  \left| \Si (\vka_0-\vka,\vka)   \right|^2 
\label{S_incoh}\\
&& {\cal C}_{\rm SFG}^{\rm (coh)}(\vka_0,\vkap_0)
=\delta(\vka_0)\delta(\vkap_0)
\left|2
\int\frac{d\vka}{(2\pi)^{3/2}}
\Theta(q_y)
H_+(\vka)H_-(-\vka)\UV(\vka)\Si(\vka, -\vka)
\right|^2
\label{S_coh}
\eeqa
\esub
\end{widetext}
where $\UV(\vka)$ and $\Si(\vka,\vkap)$ are the spectral probability amplitude defined in Eqs.(\ref{UV}) and (\ref{Sidef}).

The incoherent term (\ref{S_incoh}) originates from the up-conversion of PDC photons
which do not belong to phase-conjugated mode pairs. 
By contrast, the coherent term (\ref{S_coh}) originates from the up-conversion of photon pairs
coming from phase-conjugate PDC modes, this latter process leading to the partial reconstruction of the original coherent pump beam 
as described in the experiment in Ref.\cite{jedr2011}.
In this regard, it is convenient to introduce a special symbol for the up-conversion probability amplitude appearing in Eq.(\ref{S_coh}), namely
\bsub
\label{psipdc_inv}
\beqa
\Sicoh(\vka)&\equiv&\Si(\vka;-\vka)\\
&=&\sigma l_c'
e^{i\frac{\Deltasfg(\vka)l_c'}{2}} 
                        \sinc\frac{\Deltasfg(\vka)l_c'}{2}\;,
\eeqa
\esub
where 
\beq
\Deltasfg (\vka) \equiv \Delta (\vka,-\vka) = k_1(\vka)+k_1(-\vka)-k_0^{\rm sfg}
\label{DeltaSFGcoh}
\eeq
is the phase mismatch function for these coherent up-conversion processes. 
The function $\Sicoh$ plays a fundamental role in
determining the properties of the coherent component of the SFG field. It
describes the efficiency with which  photons belonging to a particular pair of phase-conjugated modes 
$(\vka)$ and $(-\vka)$ undergoes the inverse process of the original PDC event, i.e. 
the back-conversion into the monochromatic plane-wave mode $(\vka=0)$ corresponding to the original pump driving the PDC. 
The corresponding phase-mismatch $\Deltasfg$  
is identical to the PDC phase-matching function (\ref{deltaPW}) except for the presence of the wavenumber $k_0^{\rm sfg}$, 
which differs from the pump wave number $k_0$  only  in the case when the two crystals are not perfectly aligned
(for the type I e-oo phase-matching we are considering only the pump beam $k$-vector depends on direction). 
When the two crystals are aligned, $\Sicoh$ coincides with the probability amplitude for the reverse PDC process in the low-gain regime, 
as given by Eq.\eqref{UVlow} (apart from a constant factor). 

As can be inferred by comparing Eq.(\ref{S_coh}) and Eq.(\ref{psi_PDC}),
only the coherent contribution contains the information about the biphoton correlation $\psipdc(\vxi)$
we are looking for. 
The incoherent contribution rather acts as a background which tends to deteriorates the visibility of the correlation measurement
\cite{dayan2007}.

Notice also that the coherent component (\ref{S_coh}) factorizes into the product of the SFG field mean values:
\beq
{\cal C}_{\rm SFG}^{\rm (coh)}(\vka,\vkap)=  \langle d_0^\dagger (\vka)\rangle \, \langle d_0 (\vkap)\rangle 
\eeq
with
\beqa
\langle d_0 (\vka)\rangle&=&\delta(\vka)
\int\frac{d\vkap}{(2\pi)^{3/2}}
H_+(\vkap)H_-(-\vkap)\\
&&\hspace{2cm}\times\UV(\vkap)\Sicoh(\vkap)\nn
\eeqa

From Eqs.(\ref{sfg_correl})-(\ref{sfg_correl2}) we can evaluate
the  SFG intensity (number of photons per unit area and unit time) at the exit face of the 2nd crystal, obtaining 
\beq
I_{\rm SFG} (\vxi) \equiv  \langle
d_0^{\dagger}(\vxi)d_0(\vxi)
\rangle = 
I_{\rm SFG}^{\rm (coh)}+I_{\rm SFG}^{\rm (inc)}
\eeq
with 
\begin{widetext}
\bsub
\label{I1}
\beqa
&&I_{\rm SFG}^{\rm (inc)}
=
2\int\frac{d\vka}{(2\pi)^{3}}
 \int\frac{d\vkap}{(2\pi)^{3}}
 |V(\vka)|^2|V(\vkap)|^2 \left|\Si (\vka, \vkap) \right|^2 
\label{Iincoh1} \\
&& I_{\rm SFG}^{\rm (coh)} 
=
\left|2
\int\frac{d\vkap}{(2\pi)^{3}}
\Theta(q_y)
H_+(\vkap)H_-(-\vkap)\UV(\vkap)\Sicoh(\vkap) 
\right|^2
\label{Icoh1}
\eeqa
\esub
\end{widetext}
The intensity distribution of the SFG light in the near-field is uniform in space and time, an artifact due 
the spatio-temporal  invariance of our model deriving from the monochromatic and plane-wave pump approximation.  
By comparing Eq.(\ref{S_coh}) and Eq.(\ref{Icoh1}), we obtain the following relevant identity
\beq
{\cal C}_{\rm SFG}^{\rm (coh)}(\vka,\vkap)
= (2\pi)^3  
\delta(\vka)\delta(\vkap) I_{\rm SFG}^{\rm (coh)} 
\eeq
which shows that the coherent component of the SFG field contains
the same kind of information on the biphoton amplitude both in the space-time domain and in the spectral domain. 
The most important difference lies in that the coherent intensity in the spectral domain 
is concentrated in a single peak at the origin $(\q=0,\Omega=0)$, i.e. in the  mode corresponding to the original pump field. 
On the other hand, the  spectrum of the incoherent background is  delta-correlated in space and time, and  spreads 
over a very broad range of spatial and temporal frequencies, its particular shape being related to the phase-matching conditions inside the SFG crystal as it will be elucidated in \cite{brambilla2012}. 
This circumstance suggests that either far-field or spectral measurement
may be conveniently used to enhance the visibility of the coherent contribution with respect to the incoherent background.
\section{Retrieving the PDC correlation}
\label{sec:retrieve_psi}
Let us now investigate how the information on the correlation $\psipdc(\x,t)$ of twin beams/ twin photons  
can be effectively extracted from the coherent component of the SFG light,  given by Eq.(\ref{Icoh1}).
Making explicit the dependence on the temporal delay
and spatial shift applied to beam + and beam - contained in the transfer functions (\ref{transferH}), 
it can be written in the form
\beq
\label{IcohA}
I_{\rm SFG}^{(\rm coh)}[\Delta\vec{s},\Delta t]
=
\left|\psiout\left(\Delta \vec{s},\Delta t \right)\right|^2
\eeq
where 
\beq
\psiout(\Delta \vxi)=2
\int\frac{d\vka}{(2\pi)^3}\Theta(q_y)
e^{i\vka \cdot\Delta \vxi }
\UV(\vka)\Sicoh(\vka)\;, 
\label{psi_out}
\eeq
and $\Delta \vxi \equiv (\Delta \vec{s}, \Delta t)$. 

\begin{figure}
\centering
{\scalebox{.65}{\includegraphics*{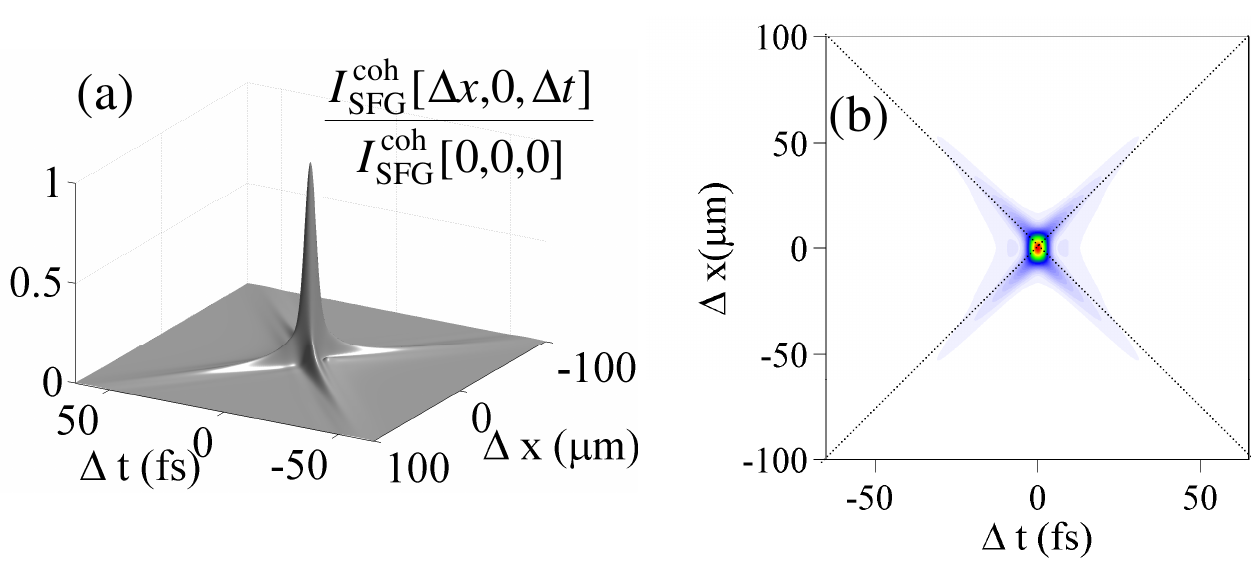}}}
\caption{(Color online) (a) Surface and (b) density plots $\Icoh[\Delta x,0,\Delta t]=|\psia(\Delta x,0,\Delta t)|^2$ normalized to its peak value,
corresponding to measurement in the configuration of Fig.\ref{fig_rotM1}a.} 
\label{figX}
\end{figure}

In this section we consider the configuration shown in Fig.\ref{fig_rotM1}a, where
mirror M1 is rotated orthogonally to the gap between the two movable mirrors so that $\Delta \vec{s}\equiv(\Delta x=2f \Delta\phi_x,0)$. 
In this case, the r.h.s. of Eq.\eqref{IcohA} coincides with the square modulus of the following quantity
\beqa
\label{psia}
&&\psia(\Delta x,0,\Delta t)\\
&&=\int\frac{d\q}{(2\pi)^2}
\int\frac{d\Omega}{2\pi}
e^{iq_x\Delta x-i\Omega \Delta t}
\UV(\q,\Omega)\Sicoh(\q,\Omega).\nn
\eeqa
The step function $\Theta(q_y)$ could be eliminated from Eq.(\ref{psi_out}) by 
exploiting the parity of integrand with respect to $q_x$, the latter being a consequence of
the radial symmetry of $\UV(\q,\Omega)$ and $\Sicoh(\q,\Omega)$ in the $(q_x,q_y)$-plane. 
The function $\psia(\Delta x,0,\Delta t)$, like $\psipdc$, is therefore an even function of both the spatial and the temporal coordinates, 
a feature not holding for the alternative configuration with mirror M1 rotating around the $x$-axis (see Fig.\ref{fig_rotM1}b) 
which will be discussed in the next section. 
Figure\,\ref{figX} plots the coherent contribution of the SFG intensity as a function of the applied spatial and
the temporal shifts. In this example,
the two crystals have the same lengths $l_c=l_c'=4\,$mm and the PDC gain is $g=8$. 
The structure preserves the nonfactorable X-shaped geometry of the original PDC correlation function $|\psipdc|^2$ shown in Fig.\ref{fig1},
with symmetric tails developing along the bisector lines $\Delta t= \pm  \sqrt{k_1'' k_1} \Delta x$ (corresponding to the diagonals of the plot frame). 
\\ 

The SFG coherent component (\ref{psia}) differs from the PDC biphoton correlation $|\psipdc|^2$ defined in Eq.(\ref{psi_PDC}) only for the presence of the SFG spectral amplitude $\Sicoh(\vka)$. The latter   
describes how the phase matching mechanism selects the spatio-temporal modes in the coherent up-conversion process.
We shall analyse in detail how it affects the PDC correlation measurement in the next subsection.
In this regard, it is useful to recast  Eq.\eqref{psi_out}  in the form 
of a convolution of the PDC biphoton amplitude $\psipdc$ - the quantity under investigation - 
with the Fourier transform of $\Theta(q_y)\Sicoh(\q,\Omega)$, i.e.
%
\beq
\label{Icoh_conv}
\psiout(\Delta\vxi\,) = \int d\vxip\psisfg(\Delta \vxi-\vxip)\psipdc(\vxip).
\eeq
where
\beqa
\label{psi_SFG} 
\psisfg(\Delta \vxi)
= \int\frac{d\vxi}{(2\pi)^3}  e^{i\vka \cdot \Delta\vxi}\Sicoh(\vka),
\eeqa
This identity has been written taking into account that $\Delta\vec{s}\equiv(\Delta x,0)$, so that the step function could be eliminated as for Eq.(\ref{psia}). Noticing that Eqs.(\ref{psi_SFG})-(\ref{psipdc_inv}) are formally identical to  Eqs.(\ref{psi_PDC})-(\ref{UVlow}), we see that  $\psisfg(\Delta \vxi)$  is proportional to the low gain biphoton amplitude in the space-time domain for the second SFG crystals. It represents the probability amplitude that a pair of photons delayed by a time $\Delta t$ and displaced by $\Delta\vec{s}$ are coherently back-converted into the plane-wave pump mode $(\q=0,\Omega=0)$.
This function can be also interpreted as the optical response function of the SFG crystal in the 
measurement of $|\psipdc|^2$ via SFG. 
 
\subsection{Thin SFG crystal limit}
\label{sec:shortSFG}
Let us assume that the two crystals are tuned for the same phase matching conditions, so that  $ \Deltasfg (\vka) = \Deltapdc (\vka)$,
but the SFG crystal is much shorter than the PDC crystal $l_c^{\prime} << l_c$. In this limit,  the PDC biphoton amplitude can be exactly reconstructed by monitoring the SFG coherent component. \\
This can be shown by inspection of Eq. \eqref{psi_out}. When the same phase matching conditions hold in the two crystals, the spectral probability amplitudes $\Sicoh (\vka)$ and $\UV (\vka)$ are peaked around the same geometrical curve $\Deltapdc (\vka)=\Deltasfg(\vka)=0$.  However, for $l_c^{\prime} << l_c$ the spectral bandwidths  in the SFG crystal are much wider than in the PDC crystal (the  bandwidths of $\Si$ exceeds those of $\UV$ by a factor $\sqrt{l_c/l_c'}\gg 1$, according to Eq.\eqref{PDC_band}. As a result, $\Sicoh (\vka)$ 
is almost constant in the region where  $\UV (\vka)$ is not negligible, close to its maximum value $\sigma l_c'$. 
Under these conditions, the filtering effect due to phase-matching in the SFG crystal becomes ineffective and we have
\beqa
I_{\rm SFG}^{(\rm coh)}[\Delta x,0,\Delta z]\approx \sigma^2 l_c'^2 |\psipdc(\Delta x,0,\Delta t)|^2
\;\;(l_c'\ll l_c)
\label{Icoh_short}
\eeqa
Thus the coherent component of the SFG output reproduces the biphoton correlation as anticipated.\\
The same conclusion can be  derived also looking at Eq. \eqref{Icoh_conv}: in the limit $l_c' <<l_c$, the biphoton amplitude $\psi_{SFG}$, defined by Eq.\,\eqref{psi_SFG}, has scales of variation in space and time much shorter than $\psipdc$, because it is the Fourier transform of a much wider spectral amplitude. It behaves therefore as a $\delta$-function inside the convolution integral \eqref{Icoh_conv}, so that the PDC biphoton amplitude
is recovered. 
\par
For completeness, we mention that in the $l_c'\rightarrow 0$ limit 
the incoherent contribution (\ref{Iincoh1}) takes the form
\beqa
&&I_{\rm SFG}^{(\rm inc)}=\sigma^2 l_c'^2 |\langle b_1^{\dagger}(\x,t)b_1(x,t)\rangle|^2\;\;\;(l_c'\rightarrow 0)
\label{Iincoh_short}
\eeqa
where
\beq
\label{PDCintensity}
\langle b_1^{\dagger}(\x,t)b_1(x,t)\rangle = \int \frac{d\q}{(2\pi)^2}\int \frac{d\Omega}{(2\pi)}|V(\q,\Omega)|^2
\eeq
is the PDC photon flux evaluated at the output face of the first crystal [as can be easily inferred from
relation (\ref{corr_PDCa})]. However, we shall see in Sec.\ref{sec:visibility} that the temporal walk-off between the
PDC field and the generated incoherent SFG field plays a fundamental role in the formation of the
incoherent component even when considering short propagation distances. 
In practice, the validity of expressions (\ref{Iincoh_short})  fails as soon as the finite length of the SFG crystal is taken into account.

\subsection{Long SFG crystal}
\label{sec:longSFG}
We now investigate how propagation in the SFG crystal affects the retrieval of the PDC correlation via the measurement of $I_{\rm SFG}^{(\rm coh)}[\Delta x,0,\Delta z]$, 
showing that both the X-shaped geometry and the strong localization of the biphoton correlation are preserved, at least when the two crystals are equally tuned. 
\begin{figure}
\centering
{\scalebox{.6}{\includegraphics*{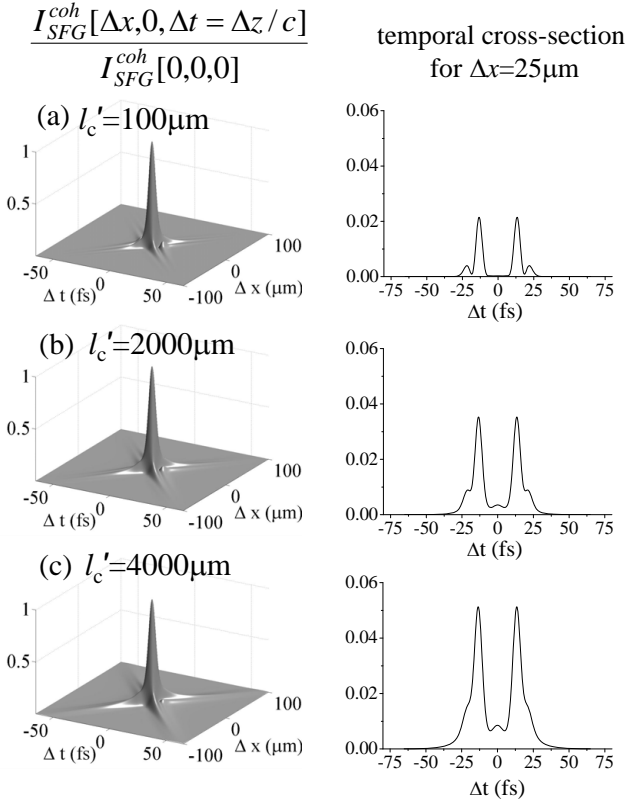}}}
\caption{Left panel: numerical evaluation from Eq.\eqref{IcohA}-\eqref{psia} of $I_{\rm SFG}^{\rm coh}[\Delta x,0,\Delta t]/I_{\rm SFG}^{\rm coh}[0,0,0]$ 
for increasing values of the SFG crystal length $l_c'$. Right panel: corresponding temporal cross section
$I_{\rm SFG}^{\rm coh}[\Delta x=25\mu m,0,\Delta t]/I_{\rm SFG}^{\rm coh}[0,0,0]$.  
The X-shaped structure is robust against propagation in the SFG crystal and the tails are enhanced with respect to the central peak as $l_c'$ increases. The PDC parameters are $g=8$, $l_c=4$mm.  
}
\label{tailsvisib}
\end{figure}
\begin{figure}
\centering
{\scalebox{0.65}{\includegraphics*{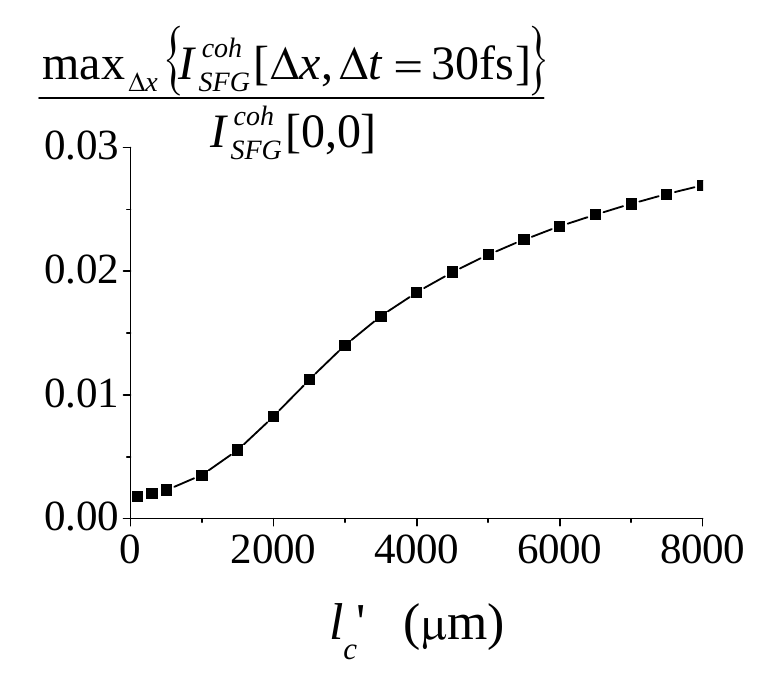}}}
\caption{Plot of the height of the tails of the retrieved correlation 
$\max_{\Delta x} \left\{ I_{\rm SFG}^{\rm coh}[\Delta x,0,\Delta t=30 {\rm fs}]\right\}$,  at a fixed time delay $\Delta t=30$fs,  
normalized to the peak value $I_{\rm SFG}^{\rm coh}[0,0,0]$, 
as a function of $l_c'$. 
}
\label{tailsdecaylog}
\end{figure}
\par
The left panels of Fig.\ref{tailsvisib} plot the retrieved quantity $I_{\rm SFG}^{\rm (coh)}[\Delta x,0,\Delta t]$ for increasing values of the SFG crystal length $l_c'$. The most evident effect is that the tails of the correlation becomes progressively more visible with respect to the central peak, as the propagation distance inside the SFG crystal increases. This is clearly shown by the temporal cross-sections at $\Delta x=25\mu$m of the same quantity (right panels of Fig.\ref{tailsvisib}). 
Figure\,\ref{tailsdecaylog} shows that the height of the tails,   at a  fixed time delay $\Delta t=30$ fs, increases almost linearly with $l_c'$ up to the value of the PDC crystal length $l_c=4$\,mm.
We expect therefore that the choice of a few millimeter long SFG crystal, with respect to that of
a very short crystal, presents a twofold advantage: {\bf i}) the total number of up-converted photons is obviously larger
and {\bf ii}) the visibility of the tails, although small with respect to that of the central peak, will be  enhanced, 
a feature which should facilitate the experimental observation of the X-shaped structure through
the scanning of $\Delta x$ and $\Delta t$.
A further advantage lies in that the overall visibility of the coherent component
with respect to the incoherent background improves substantially as the SFG crystal length is raised
above a few hundred of microns, as it will be shown in Sec.\ref{sec:visibility}.

\begin{figure}
\centering
{\scalebox{0.46}{\includegraphics*{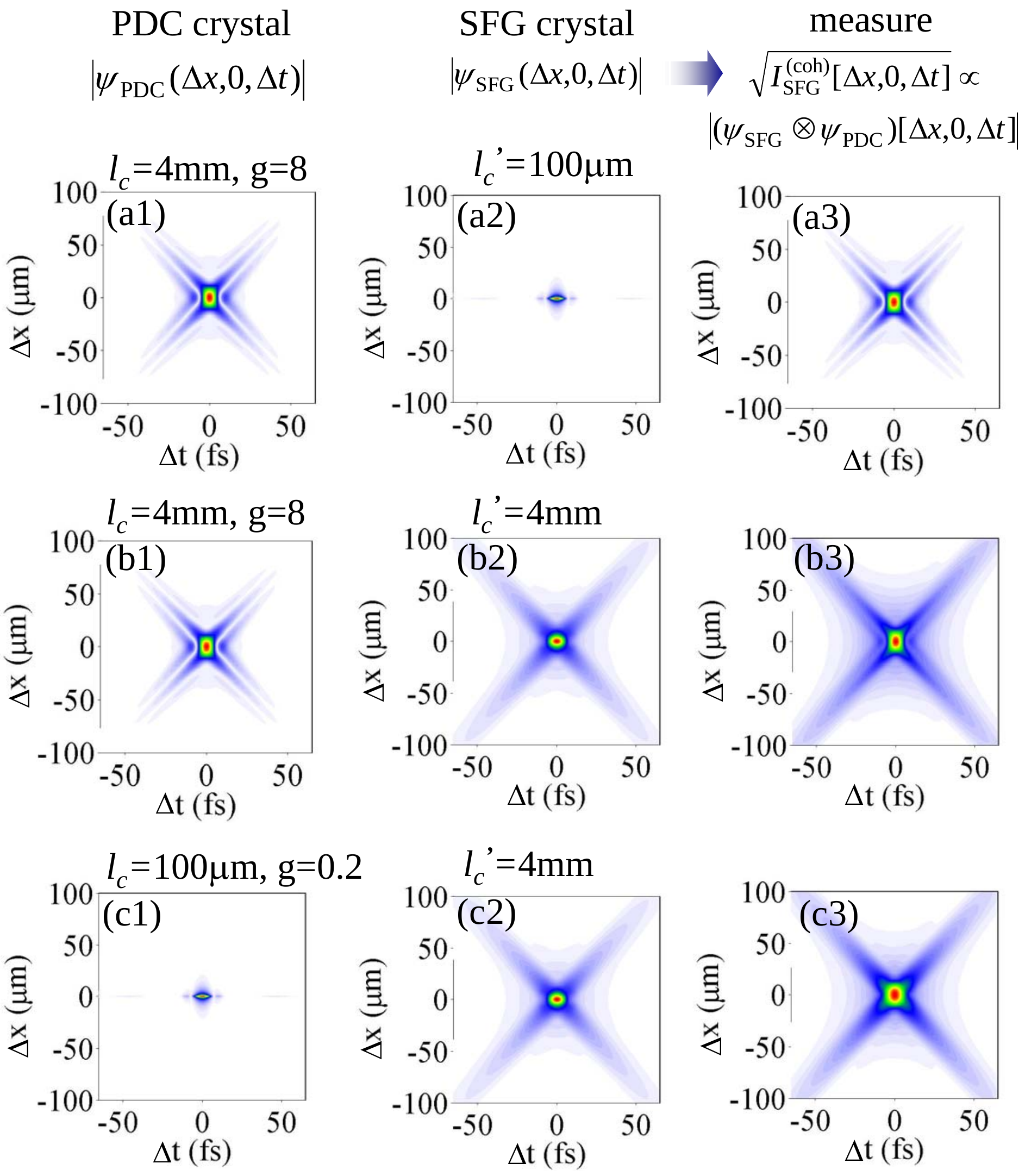}}}
\caption{(Color online) The left and central panels shows the density plots of $|\psi^{\rm PDC}(\Delta x,0,\Delta t)|$ and $|\psi^{\rm SFG}(\Delta x,0,\Delta t)|$, respectively, 
considering either short or long crystals (the values of $l_c$ and $l_c'$ are indicated). 
The square root of the measured quantity, $\sqrt{I_{SFG}^{coh}}=|\psiout(\Delta x,0,\Delta t)|$, obtained by convolving 
$\psipdc$ and $\psisfg$, is shown in the right panel. See text for a discussion. 
}
\label{fig_SFGresponse}
\end{figure}
\par 
An explanation of the behaviour depicted in Fig.\ref{tailsvisib} can be obtained considering the convolution integral in Eq.\eqref{Icoh_conv} and referring to the density plots of Fig.  \ref{fig_SFGresponse}, which illustrate  the effect of the convolution (\ref{Icoh_conv}) under three different 
conditions: \\ 
{\bf (a)}  $l_c'\ll l_c$ : the SFG crystal response function $\psisfg$ behaves as a Dirac $\delta$-function both in the spatial and temporal 
domain (see plot a2), so that the 2nd crystal works as an ultrafast correlator with instantaneous and localized response. 
The coherent component of the SFG intensity (plot a3) provides therefore the PDC biphoton amplitude 
square modulus $|\psipdc|^2$, in agreement with the short crystal limit result (\ref{Icoh_short}).\\
{\bf (b)}   $l_c \sim l_c'$ : if the SFG crystal has a length comparable to that of the PDC
crystal, it displays a nonlocal spatio-temporal response. However, provided the two crystals are tuned for the same phase matching conditions, the response of the SFG crystal has the same geometrical X-shape as the original biphoton correlation [see plots (b1) and (b2)]. 
This does not come unexpected: as a matter of fact, the function $\psisfg$ is identical to $\psipdc$ in the low PDC gain limit (apart from a constant factor). In the high gain regime of PDC considered in this example, the PDC biphoton amplitude displays a faster decay of the tails than its low gain counterpart,\footnote{The faster decay of the PDC correlation $\psipdc$ along its tails for high gain is due to the broadening of the spectral amplitude with $g$ described by Eq.(\ref{PDC_band_g}).} 
but nevertheless keeps the same geometrical shape.  Therefore $\psiout$ is  
the convolution of two functions which have almost the same  X-shaped structure in the space-time domain, and the
main effect of this convolution is to enhance the weight of the tails with respect to the central peak,
as can be seen from plot (b3). 
Most-importantly from our point of view, the non-factorable X-shaped geometry of the PDC biphoton amplitude is preserved, as well as its strong localization in time (or in space) when particular temporal (or spatial) cross-sections are considered. \\
{\bf (c)}   $l_c \ll l'_c$.
Because of the symmetrical role played by $\psipdc$ and $\psisfg$ in Eq.(\ref{Icoh_conv}),
a configuration with $l_c\ll l_c'$ would provide a direct measure of $|\psisfg|^2$, which is proportional to the 
PWP biphoton amplitude in the SFG crystal at low parametric gains (accordingly, the density plots (c2) and (c3) are nearly identical,
as for the plots (a1) and (a3) in case a). 
However, in this case the SFG photon flux would be strongly reduced because of the lower gain of the PDC field associated with short crystals.

\section{Alternative detection scheme}
We now focus on the alternative configuration (b) for mirror M1 shown in Fig.\ref{fig_rotM1}b, considering rotations performed
around the $x$-axis. In contrast to configuration (a) treated in the previous section, the step function $\Theta(q_y)$ cannot be eliminated from Eq.(\ref{psi_out}) and the SFG coherent component is now given by the square modulus of
\beqa
&&\psib(0,\Delta y,\Delta t)
=2
\int\frac{d\q}{(2\pi)^2}
\int\frac{d\Omega}{2\pi}
e^{iq_y\Delta y-i\Omega \Delta t}
\Theta(q_y)\nn\\
&&\hspace{3cm}\times\UV(\q,\Omega)\Sicoh(\q,\Omega)
\label{psib}
\eeqa
with $\Delta y=2f \Delta\phi_y$.

\begin{figure}[h]
\centering
{\scalebox{.65}{\includegraphics*{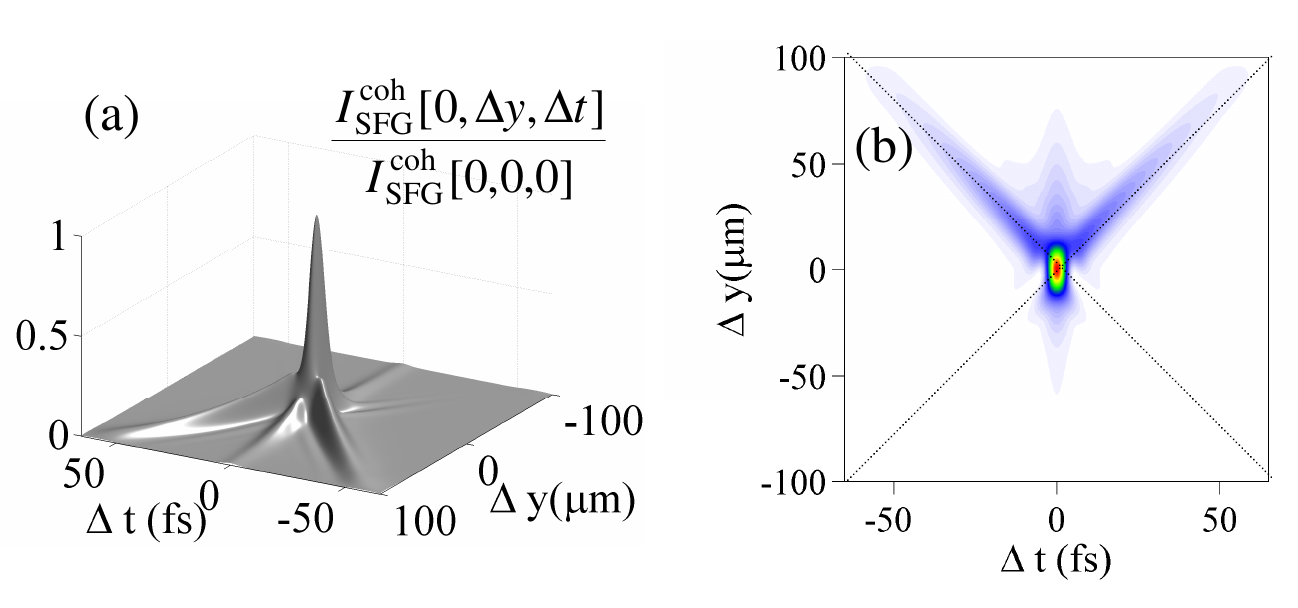}}}
\caption{(Color online) (a) Surface and (b) density plots $\Icoh[0,\Delta y,\Delta t]=|\psia(0,\Delta y,\Delta t)|^2$ normalized to its peak value,
corresponding a measurement in the configuration of Fig.\ref{fig_rotM1}a.} 
\label{figXvsV}
\end{figure}
Because of the presence of the step function, the even symmetry with respect to the spatial coordinate is lost.
As shown in Fig.\ref{figXvsV}, the retrieved correlation function in the $(\Delta t,\Delta y)$-plane display a V-shaped geometry:
the tails extends along the lines $\Delta t= \pm  \sqrt{k_1'' k_1} \Delta y$ in the $\Delta y>0$ half-plane, while they
disappears for $\Delta y<0$. Moreover, compared to the result obtained in configuration (a) shown in Fig.\ref{figX}, the tails are 
strongly enhanced with respect to the central peak. From simple geometrical optics considerations, it can be shown
that negative values of the rotation angle $\Delta \phi_y$ of mirror M1 around the $x$-axis prevents the overlap of beam + and beam - inside
the SFG crystal, so that the up-conversion efficiency becomes much lower under this condition (i.e for $\Delta y<0$). On the contrary,
a negative rotation $\Delta \phi_y$ increases the overlap of beam + and beam -, enhancing thereby the up-conversion efficiency compared to configuration (a) in the $\Delta y>0$ region.
\begin{figure}
\centering
{\scalebox{.69}{\includegraphics*{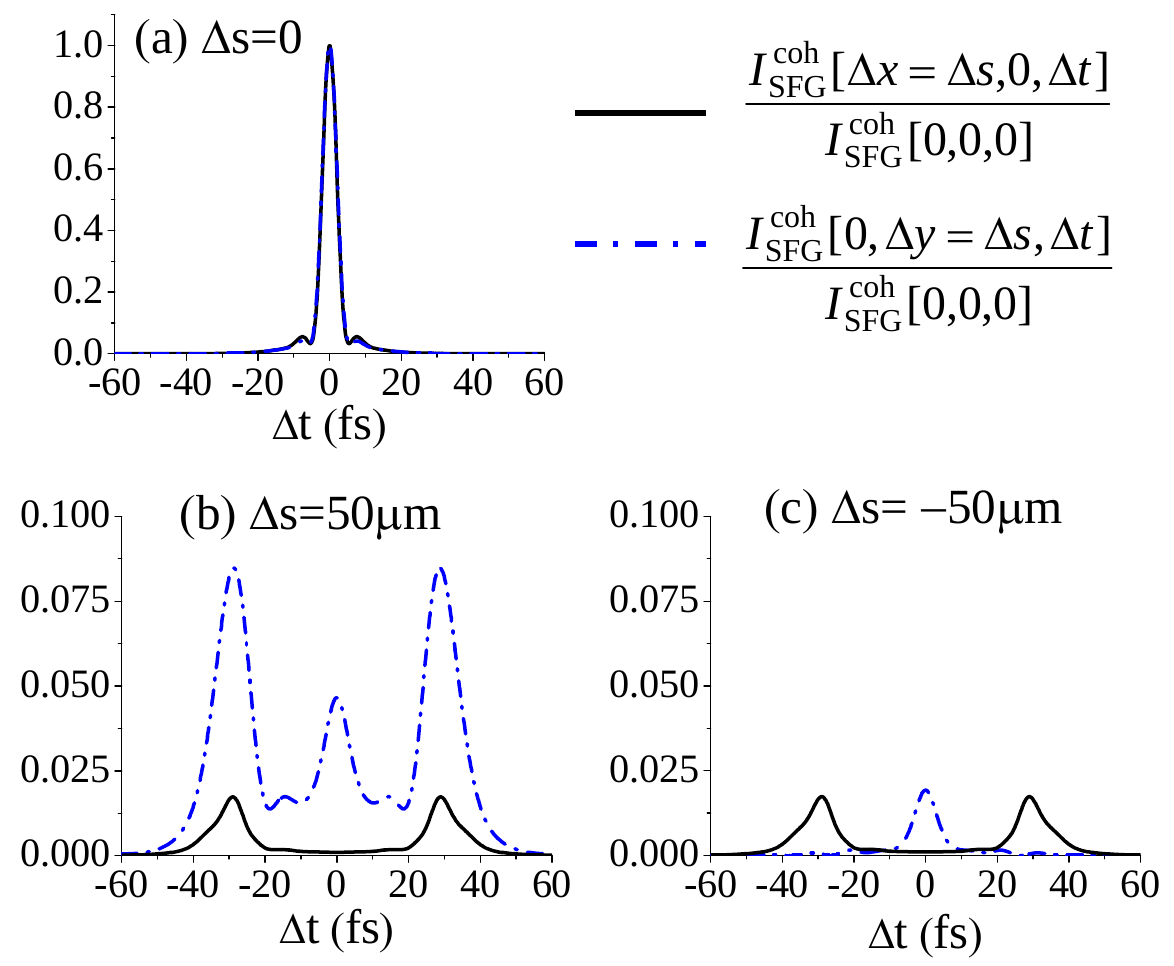}}}
\caption{(Color online) Temporal cross-sections of the SFG coherent component obtained by varying $\Delta t$   
for (a) $\Delta s=0$, (b) $\Delta s=50\mu$m and (c) $\Delta s=-50\mu$m. The solid line corresponds to configuration (a), 
the dotted dashed line corresponds to configuration (b).
} 
\label{figXvsVsections}
\end{figure}

Noticing that the complex amplitude $\psia(\Delta x,0,\Delta t)$ and $\psia(0,\Delta y,\Delta t)$ given by Eq.(\ref{psia}) and Eq.(\ref{psib})
are linked through the relation
\beqa
\label{psisym}
&&\psib(\Delta x,0,\Delta t)=\frac{1}{2}\left[\psia(0,\Delta y=\Delta x,\Delta t)\right.\nn\\
&&\hspace{2cm}\left.+\psia(0,\Delta y=-\Delta x,\Delta t)\right],
\eeqa
from which it can be inferred that the temporal profile across the central peak [obtained by setting $\Delta x=0$ in case (a), $\Delta y=0$ in case (b)]
is identical in the two configurations (see Fig.\ref{figXvsVsections}a).
On the other hand, being $\psib(0,\Delta y,\Delta t)\approx 0$ in the $\Delta y<0$ region not too close to the central peak,
relation (\ref{psisym}) implies that $\Icoh[0,\Delta y,\Delta t]\approx 0$ for $\Delta y<0$ and $\Icoh[0,\Delta y,\Delta t]\approx 4\times 
\Icoh[\Delta x=\Delta y,0,\Delta]$ 
for $\Delta y>0$.  The tails of the coherent SFG component, extending only in the $\Delta y>0$ region in configuration (b), are thus about
four times more intense than in configuration (a). This behaviour is shown in Fig.\ref{figXvsVsections}b,c, which compares the temporal profile in 
the two configurations for the spatial shifts $\Delta s=50\mu$m and $\Delta s=-50\mu$m respectively.

We notice that configuration (b) offers the relevant benefit that the tails are more visible and can be therefore detected more easily
than in configuration (a). Although in principle configuration (b) does not allow the reconstruction of the symmetric X-shaped correlation as
configuration (a), we found the following approximate empirical relation that links the results in the two configurations:
\beqa
\Icoh[\Delta x=\Delta y,0,\Delta t]\\ 
&&\hspace{-3.7cm}\approx\frac{1}{4}\left[\sqrt{\Icoh[0, \Delta y, \Delta t]}
-\sqrt{\Icoh[0,-\Delta y,\Delta t]}\right]^2\,\text{for}\,\Delta y\neq 0,\nn
\eeqa
this approximation holding as long as $\Delta y$ is sufficiently large, i.e. far from the central peak.
This relation should allow to infer from experimental data collected in configuration (b) the data that would be collected in
configuration (a) , thus allowing the reconstruction of the X-shaped PDC correlation in configuration (b) .

\section{Visibility of the information}
\label{sec:visibility}
The issue of the visibility of the information contained in the coherent SFG contribution,
against the incoherent background, is a crucial one, especially for the observation of the tails of the PDC X-shaped correlation.

As already remarked, the expression \eqref{sfg_correl2}  of the spectral coherence function of the SFG light, suggests that the visibility should be greatly enhanced by measuring the SFG light in the far-field of the second crystal, where  the coherent contribution propagating in the forward direction is separated from the incoherent background propagating over a broad angle. This issue will be discussed in Sec.\ref{sec:farfield}. We start here by considering  the visibility of a measurement of the total number of SFG photons, which provides a useful estimation of the overall weight
of the coherent component with respect to the incoherent one.
\subsection{Bucket detection of the  SFG photons} 
\label{sec:bucket} 
The photon fluxes $I_{\rm SFG}^{\rm (inc)}$ and $I_{\rm SFG}^{\rm (coh)}[\Delta x,0,\Delta t]$ given in Sec.\ref{sec:SFGfield}, Eqs.(\ref{I1}), have been evaluated at the SFG crystal output face and are are uniform in the transverse plane because of the artifact of the plane-wave pump approximation. Within this limit, the ratio
\beq
\label{visibility}
{\cal V}=\frac{I_{\rm SFG}^{\rm (coh)}[0,0,\Delta z=0]}
{I_{\rm SFG}^{\rm (inc)}+I_{\rm SFG}^{\rm (coh)}[0,0,\Delta z=0]}
\eeq
defines therefore the visibility of the correlation peak at $\Delta x=\Delta t=0$ against the incoherent background
either in the case of a near-field detection of the SFG intensity or assuming all the SFG photons are collected without
discrimination (e.g. by using a bucket detector).  

\begin{figure}[H]
\centering
{\scalebox{.65}{\includegraphics*{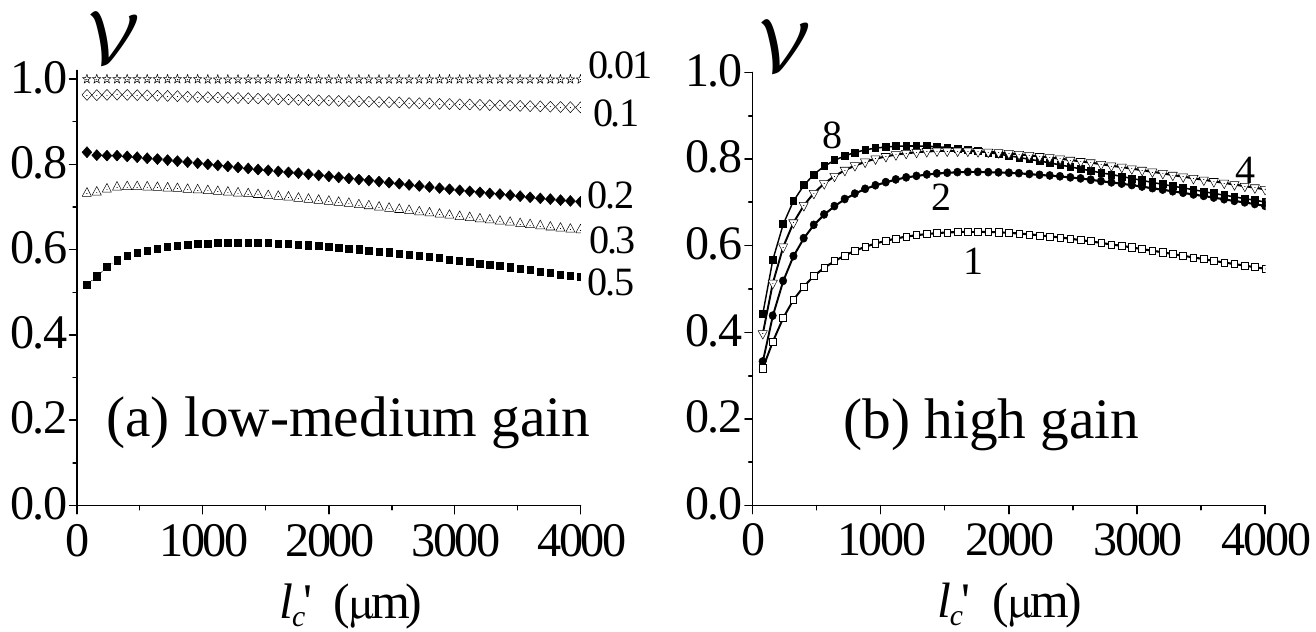}}}
\caption{Visibility of the coherent component with respect to the incoherent background as a function of the length of the
SFG crystal $l_c'$, for (a) medium-low and (b) high parametric gain values (indicated in the figure). The length of the PDC crystal, $l_c=4$mm, is
kept fixed. The selected temporal bandwidth extends from 850\,nm up to 1400\,mm. 
}
\label{fig_visib}
\end{figure}

Figure \ref{fig_visib} plots the visibility ${\cal V}$ as a function of the length of the SFG crystal $l_c'$
for increasing PDC parametric gain, starting from values of $g$ below unity (Fig.\ref{fig_visib}a) up to very high values (Fig.\ref{fig_visib}b), the latter corresponding to the regime of the experiment reported in \cite{jedr2011}.  
The contributions of both the coherent and the incoherent components (\ref{Icoh1}) and (\ref{Iincoh1}) have been
estimated numerically with a Monte Carlo integration, assuming a 550\,nm FWHM temporal bandwidth is selected with a super-Gaussian frequency filter from 850 nm up to 1400 nm. 

In a regime of low parametric gain, i.e. for $g\ll 1$, the PDC spectral probability amplitude
$\UV(\vka)$ scales as $g$ [see Equation \eqref{UVlow}], while the spectrum $|V(\vka)|^2$  given by Eq.\eqref{V} scales as $g^2$ . 
For this reason $I_{\rm SFG}^{\rm (coh)}$ and $I_{\rm SFG}^{\rm (inc)}$ scale as $g^2$ and $g^4$,respectively, 
and the visibility (\ref{visibility}) is close to $100\%$. This behaviour can be physically understood by noticing
that the probability of finding pair of photons that are not twins becomes exceedingly low in this regime.\\
As the PDC parametric gain increases, the incoherent component rapidly increases  and becomes comparable to the coherent one. 
As a consequence,  ${\cal V}$ decreases to lower values, the degradation being particularly relevant in 
the $l_c'\ll l_c$ limit. For longer propagation distances in the SFG crystal, the generation of incoherent SFG photons
becomes less efficient because of spatial walk-off and GVM, while the coherent component is not affected by those phenomena,
as it will be further discussed in \cite{brambilla2012}. 
The visibility remains therefore above $60\%$ even for high PDC gains, 
as long as the propagation distance inside the SFG crystal exceeds a few hundred micrometers. However, we notice that such 
visibility, although not negligible, would make the detection of the full X-correlation very challenging in a practical implementation, because the tails would be hidden by the incoherent background. 
\begin{figure}
\centering
{\scalebox{.8}{\includegraphics*{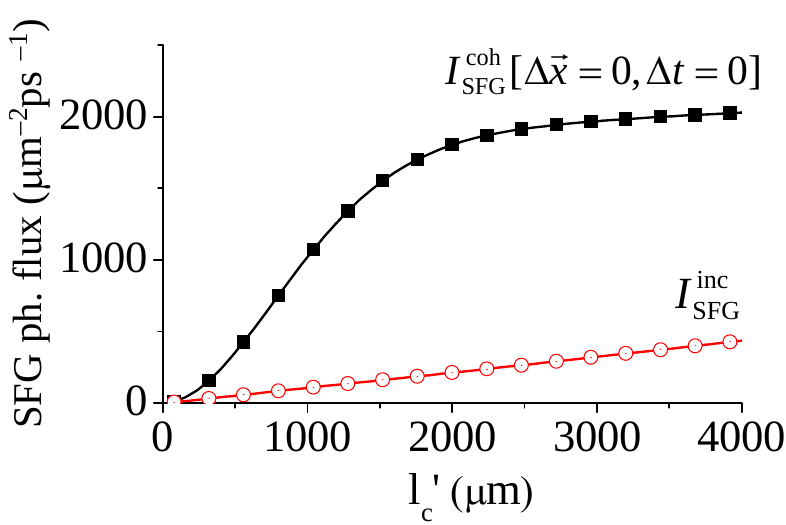}}}
\caption{Estimation of the coherent and incoherent SFG photon fluxes, $I_{\rm SFG}^{\rm (inc)}$ and $I_{\rm SFG}^{\rm (coh)}[0,0,0]$, as a function of the SFG crystal length $l_c'$. The PDC crystal parameters are $g=8$ and $l_c=4000\mu$m. 
}
\label{fig_visib2}
\end{figure}
\par
Fig.\ref{fig_visib2} displays separately $I_{\rm SFG}^{\rm (coh)}[0,0,0]$ and $I_{\rm SFG}^{\rm (inc)}$ as a function of $l_c'$,  evaluated  in the high-gain case  ($g=8$) from Eq.(\ref{I1}). 
It shows that the coherent component increases almost quadratically up to $l_c'\sim 1$\,mm, while for larger propagation distance in the SFG crystal the coherent up-conversion becomes less efficient. The initial quadratic behaviour reflects the result of  Eq. \eqref{Icoh_short} derived in
the "thin SFG crystal" limit, where the filtering effect due to the SFG spectral probability amplitude $\Sicoh(\q,\Omega)$ 
in Eq.(\ref{Icoh1}) is almost ineffective. 

By inspecting the expression of the typical bandwidths of $\Sicoh(\q,\Omega)$ and $\UV(\q,\Omega)$ [only the latter depends 
on the parametric gain according to Eq.\eqref{PDC_band_g}], we find that they become comparable 
for $l_c'=l_c\sqrt{\pi^2/(\pi^2+g^2)}\approx 1.5$\,mm for $g=8$, $l_c=4$\,mm.  
For $l_c'$ above this characteristic length the SFG spectral probability amplitude
$\Sicoh(\q,\Omega)$  becomes narrower than $\UV(\q,\Omega)$, and the integral \eqref{Icoh1} spans a volume in the Fourier space 
$\propto l_c^{\prime \, -3/2} $, so that $I_{\rm SFG}^{\rm (coh)}$ does not scale any more as  $l_c'^2$. 
\par
On the other side, the  SFG incoherent component increases only linearly with $l_c'$,
except for very short propagation distances for which approximation (\ref{Iincoh_short}) holds. This  is due to the fact 
that  the up-conversion probability $|\Si(\vka,\vkap)|^2$ appearing inside the integral at r.h.s. of \eqref{Iincoh1} decays   rapidly  when $\vkap \ne -\vka$, because of the effect of  the  spatial  walk-off and the temporal group velocity mismatch  
 arising between the ordinary fundamental beam and the extraordinary  up-converted beam. This issues will be further elucidated in a forthcoming related publication \cite{brambilla2012}. 
\subsection{Far-field detection of the SFG light} 
\label{sec:farfield} 
\begin{figure}
\centering
{\scalebox{.6}{\includegraphics*{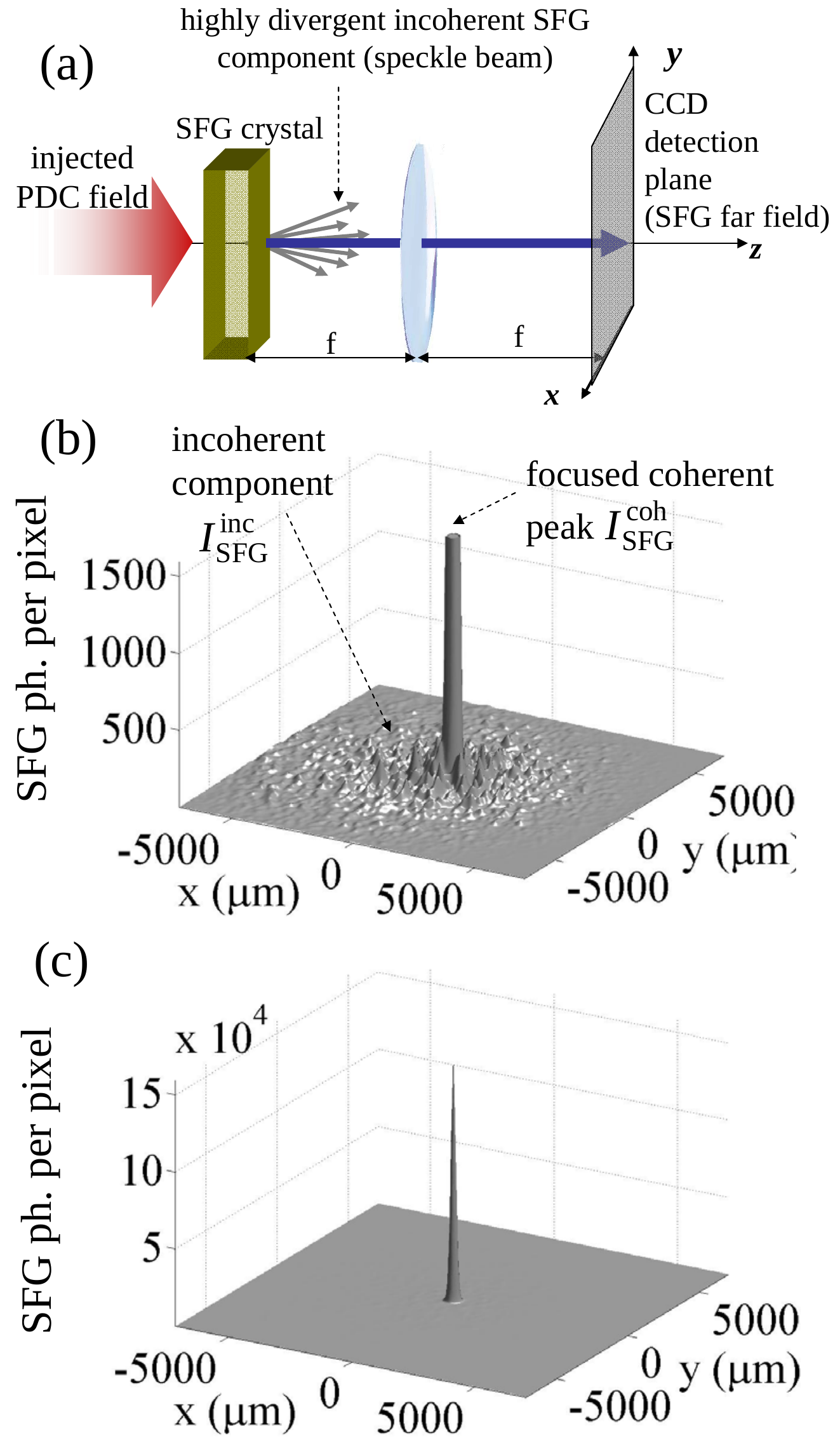}}}
\caption{(Color online) Numerical 3D+1 modeling of the experiment. (a) Detection scheme with a 2-f lens system
to observe the SFG far field (f=20cm). (b) Far-field distribution of the SFG light for $\Delta \x=\Delta t=0$: the narrow central peak (truncated to $1\%$ of its peak value) is the coherent component, arising from the up-conversion of phase conjugated photons.  The incoherent contribution gives rise to the broad speckled distribution. (c) Full scale plot, showing that the visibility of the coherent component is close to $100 \%$. g=8,  $w_p=600\mu$m, $\tau_p=1$ps. The $z$-axis scale gives the estimated number of photons on the $56{\rm \mu m}\times56 {\rm \mu m}$ pixels of the numerical grid. 
} 
\label{fig:farfield}
\end{figure}

According to the plane-wave pump result of (\ref{sfg_correl2}), the spectral distribution of the PDC light consists of a coherent contribution, concentrated in the original plane-wave pump  mode at $\q=0$ and $\Omega=0$, superimposed to an incoherent background that spreads over a large bandwidth of spatio-temporal modes. The PWP approximations leads to artificial divergences (the $\delta$-functions factors) that do not allow
a direct evaluation of the visibility of the coherent component against the incoherent background in the Fourier domain. 
For this reason, we implemented a full 3D+1 numerical simulation of our proposed setup (see Fig.\ref{fig1}), and modeled the result of the detection of the SFG intensity distribution in the far field of the SFG crystal, as shown in fig.\ref{fig:farfield}a. 
The generation of the broadband PDC field is simulated in the framework of the Wigner
representation as described in \cite{brambilla2004}, taking into account both the finite cross section $w_p$ and duration $\tau_p$ of the pump pulse,  and the phase-matching conditions inside the two crystals (the full BBO Sellmeier dispersion relation \cite{boeuf2000} are used in the simulation). 
The propagation in the two crystals, described by Eqs.(\ref{propPDC}) and (\ref{propSHG}),
is simulated through a pseudo-spectral (split-step) method using a $256\times256\times512$ numerical grid in the 
$(\x,t)-$ and the $(\q,t)-$spaces.
The parameters of the numerical simulations are chosen to reproduce 
the conditions of the experiment being developed in Como  \cite{jedr2011},  
which operates in a pulsed regime of high parametric gain $(g\approx 8)$.  
The broadband PDC field injected into the SFG crystal after the extraction of the pump beam undergoes the three-wave mixing process described
by Eqs.(\ref{propSHG}) and the up-converted SFG field is mapped into the far-field with an f-f lens system, 
as shown schematically in Fig.\ref{fig:farfield}a.
We expect that a single stochastic realization of our simulations roughly reproduces the field distribution obtained from 
a single pump shot. 
\par
The typical far-field intensity distribution of the up-converted SFG field obtained in the detection plane from a Gaussian pump pulse of waist
$w_p=600\mu$m and duration $\tau_p=1\,$ps is shown in Fig.\ref{fig:farfield}b,c. 
We verified that for the chosen PDC gain g=8
the injected PDC field (not shown in the figure) is only slightly depleted during propagation in the second crystal ($< 0.1\%$),
a feature which confirms the validity of the perturbative approximation (\ref{inputoutput_sfg}) for the SFG field used in the PWPA model. 
The narrow central peak results from coherent processes in which pairs of phase conjugate photons back-convert to a coherent field component reproducing partially the far-field distribution of the original pump beam. This peak is fixed, in the sense that it is reproduced identically in each stochastic simulation. The broad speckled background instead is noisy and changes in each realization, giving rise on average to a broad distribution. It originates from the incoherent processes where not phase-conjugated photon pairs are up-converted. 
These kind of simulations reproduce very closely the results on the far-field detection of SFG reported in \cite{jedr2011} (see in particular Fig.4 therein). \\
The full scale plot in Fig.\ref{fig:farfield}\,c shows that in such ideal conditions (perfect imaging, no dispersive optical elements, no misalignments between the two crystals) the visibility of the coherent vs incoherent component for $\Delta x=\Delta t=0$ is very high, close to $100\%$. 
We can therefore expect that this far field detection scheme is well suited for performing the measurement of the full X correlation.

\begin{figure}
\centering
{\scalebox{.75}{\includegraphics*{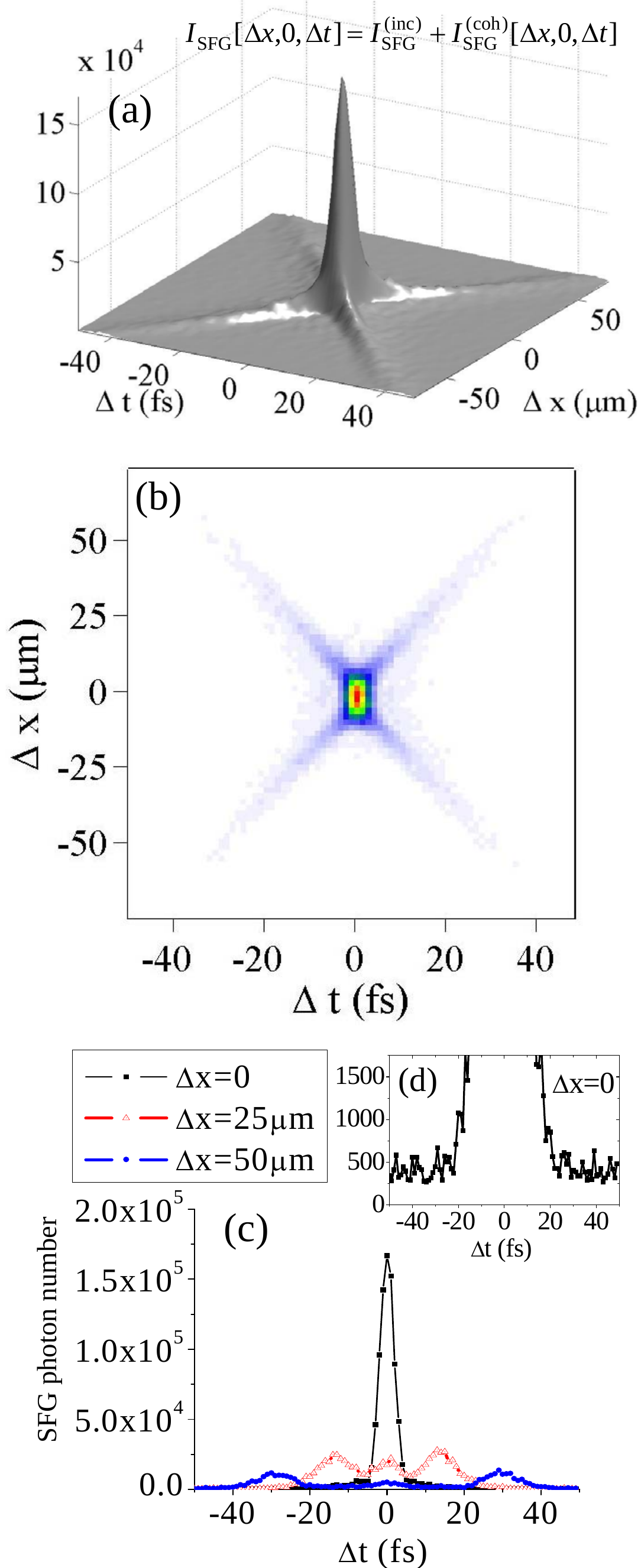}}}
\caption{(color online) Simulation of the reconstruction of the X-correlation via SFG under ideal conditions based on the far-field detection scheme of Fig.\ref{fig:farfield}\,a.  (a) Surface and (b) density plots of the number of SFG photons measured in the central pixel (at $\x=0$), 
obtained by scanning $\Delta x$ and $\Delta t$ on a $80\times80$ grid. (b) Temporal profiles for three different values of $\Delta x$ showing the transition from the central peak to the double peak profile. 
The inset (d) shows the baseline of the $\Delta x=0$ profile,  corresponding to the incoherent background with around 350 photons per pixel, 
about $0.2\%$ of the peak value at $\Delta x=\Delta t=0$. $g=8$,  $w_p=600\mu$m, $\tau_p=1$ps. 
}  
\label{fig:X}
\end{figure}
Figure \ref{fig:X} simulates the reconstruction of the X-shaped correlation obtained by monitoring the number of photons in the central peak of the SFG far-field as a function of the temporal delay $\Delta t $ and the spatial shift $\Delta x$ imposed on the two twin PDC components  (more precisely, it reports the number of SFG photons over a pixel 56\,${\rm \mu m}$ x 56\,${\rm \mu m}$  wide,  in the focal plane of a lens with f=20 cm, placed at a focal distance from the SFG crystal). We verified that for the chosen pump pulse parameters, the retrieved structure in the $(\Delta t,\Delta x)$-plane is very close to that obtained by evaluating the coherent component (\ref{psia}) within the PWPA (see e.g. Fig.\ref{figXvsV}a).
Notice that despite the fact that both the coherent and the incoherent components are displayed in Fig.\ref{fig:X}, the visibility of the tails of the correlation with respect to the speckled background is very high.  
The relative weight of the incoherent background, about 0.2\% of the coherent peak value, can be inferred from the height of the baseline of the temporal profile plotted in inset (d) for $\Delta x=0$. 

\section{Fragility of the correlation measurement}
\label{sec:fragility}
The previous simulations that display nearly 100\% visbility for the correlation measurement have been obtained 
assuming a perfect imaging of the PDC source into the SFG crystal input face. 
The coherent component $\propto |\psiout(\Delta x,0,\Delta t)|^2$ is however strongly phase-sensitive and
therefore extremely fragile against dispersion,
imperfect imaging conditions, as well as small misalignments between the two crystals.
For this reason, the slightest imperfection in the imaging device that maps the PDC output plane into the SFG input plane
deteriorates the coherent component of the SFG field which contains the correlation information. 
On the other hand, the incoherent component (\ref{Iincoh1}) remains  unaffected
because of its phase-insensitive nature, so that
the overall visibility of $|\psiout(\Delta x,0,\Delta t)|^2$ is strongly sensitive to those imperfections. 
In the following we shall study the effects of the main sources of experimental imperfection.
\subsection{Temporal dispersion} 
\begin{figure}[h]
\centering
{\scalebox{.49}{\includegraphics*{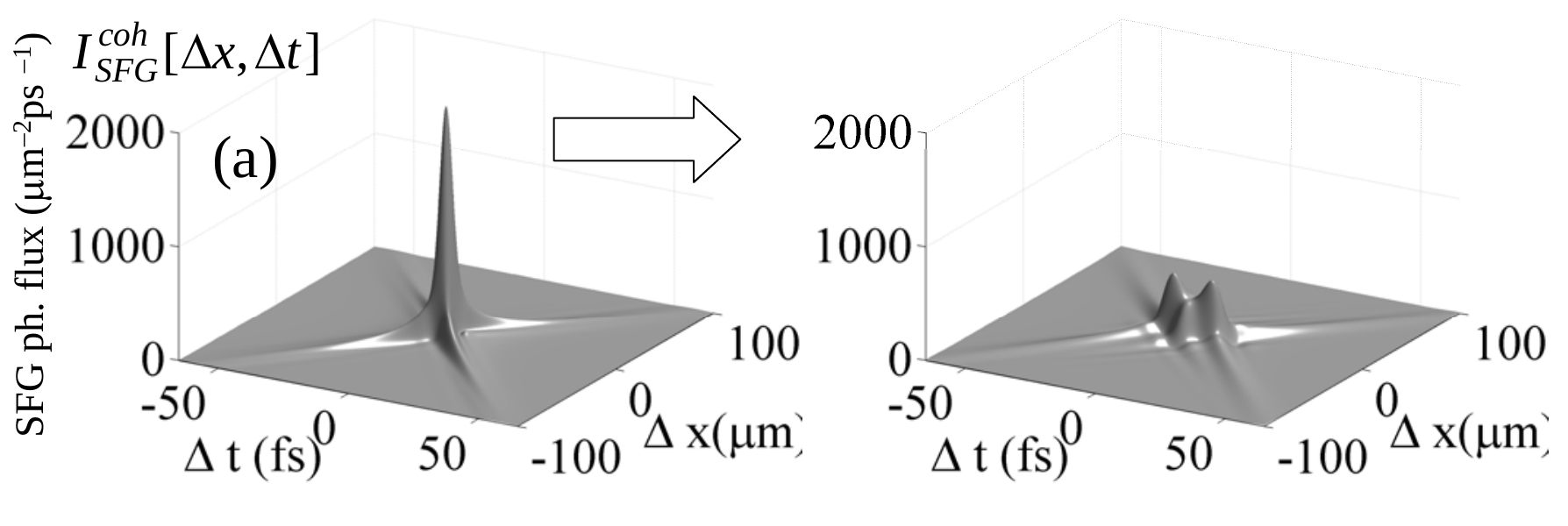}}}
\caption{Simulation of the effect of a dispersive element between the PDC and SFG crystal obtained using the PWPA model. 
Spatio temporal correlation $|\psiout|^2$.  reconstructed (a) in ideal conditions, (b) in the presence of a 0.5\,mm slab of BK7 glass between the PDC and the SFG crystals. 
}
\label{fig_dispersion}
\end{figure}

It is well known \cite{dayan2005, peer2005b, dayan2007, odonnel2009} that the scheme is very sensitive to the presence of any dispersive optical elements. As an example, 
Fig.\,\ref{fig_dispersion} shows the effect of  a half millimeter thick slab of BK7 glass inserted in the propagation path between the PDC and the SFG crystals.  
It shows that not only the central correlation peak is broadened, but the X-shaped structure of correlation is strongly distorted. 
As a matter of fact, in order to  compensate the detrimental effect of dispersion introduced by optical lenses, prisms have been used in the experimental works \cite{dayan2005, peer2005b,  odonnel2009}; a valid alternative approach, which we implemented in \cite{jedr2011,jedr2012} is rather to replace dispersive lenses 
by achromatic  parabolic mirrors. 
\subsection{Imperfect imaging} 
Because of the smallness of the variation scale of the PDC  X-correlation along the spatial dimension ($\lesssim 10 \mu$m), 
we expect that the scheme shows very little tolerance with respect to errors in the positioning of 
the SFG crystal with respect to the image plane of the telescopic system illustrated in Fig.\ref{fig2}.
Under ideal imaging conditions, the spatial width of the correlation peak is indeed determined
by the collected PDC bandwidth considered in the simulation, i.e. $2q_{\rm max}\approx 0.6\mu{\rm m}^{-1}$, 
which gives a spatial correlation profile of width $\Delta x_{\rm fwhm}\approx 11\,\mu$m (see Fig.\ref{figPM}a and Fig.\ref{fig1}c). 
\par
Let us suppose for example that the second crystal entrance face is set at a distance $f+\Delta z_{\rm img}$ 
from the second lens of the 4-f system.
This error introduces an additional phase factor 
to the transfer function product that enters in Eq.(\ref{Icoh1}), which is now given by
\beq
\label{phase_dz_img}
H_+(\q,\Omega)H_-(-\q,-\Omega)      
=e^{-i\Omega \Delta t}e^{iq_x \Delta x}
e^{-i\frac{c q^2 }{\omega_1(1-\Omega^2/\omega_1^2)}\Delta z_{\rm img}}
\eeq 
The condition for neglecting the diffraction term reads
\beq
\Delta z_{\rm img}\ll z_{\rm DOF}\equiv
\frac{\pi^2}{ \lambda q_{\rm max}^2}
\left(
1-\frac{\Omega_{\rm max}^2}{\omega_1^2}
\right)
\eeq  
where $q_{\rm max}$ is the maximal transverse wave-vector of the PDC emission in the collected bandwidth. 
For the temporal bandwidth $2\Omega_{\rm max}=0.965\times10^{15}$Hz assumed in the simulations,
we have a tolerance on $\Delta z_{\rm img}$ on the order of hundred micrometers.

\begin{figure}
\centering
{\scalebox{.7}{\includegraphics*{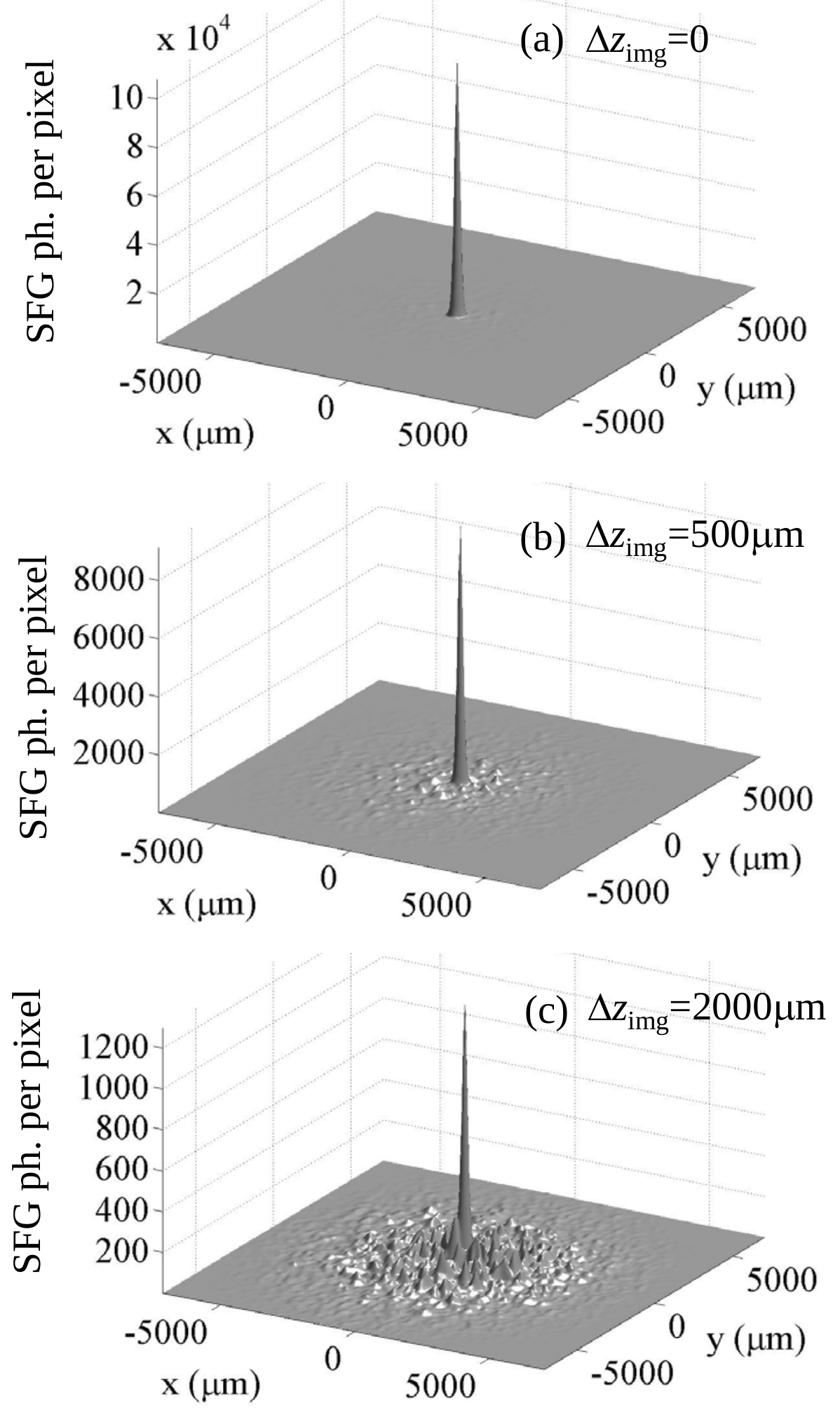}}}  
\caption{Simulation of the far-field distribution of the SFG light evaluated for increasing values of the error in the imaging plane positioning $\Delta z_{\rm img}$.}
\label{fig_dzimg}
\end{figure}

This is  confirmed by our numerical modeling of the experiment. 
A first effect of an error in the imaging system is that the efficiency of the coherent up-conversion drops substantially with respect to
the ideal configuration.  Figure \ref{fig_dzimg} 
shows the far-field distribution of the SFG light (as it would be detected in the scheme of Fig.\ref{fig:farfield}a), for increasing values of the error in the imaging plane $\Delta z_{\rm img}$. A displacement of  $500\,{\rm \mu m}$ is sufficient to decrease the back-conversion efficiency of the twin photons by a factor 12, and for $\Delta z_{\rm mm}=2\,$mm the coherent peak becomes comparable in magnitude to the incoherent background (which is not sensitive to $\Delta z_{\rm img}$). 
\par
Even more dramatical is the deterioration of the shape of the correlation function. 
\begin{figure*}
\centering
{\scalebox{1.}{\includegraphics*{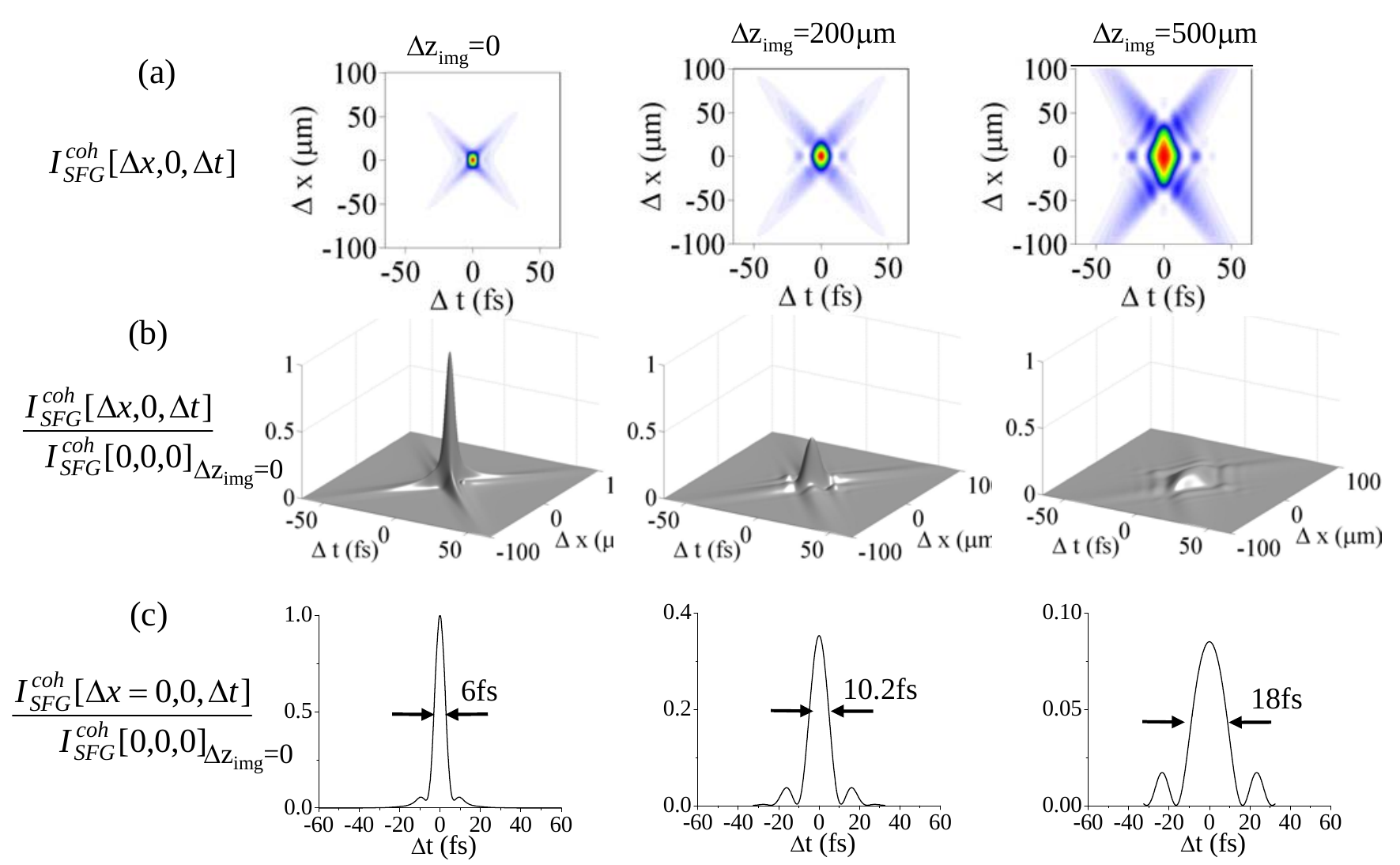}}}
\caption{(color online) (a) Behaviour of the reconstructed correlation $|\psiout(\Delta x,0,\Delta t)|^2$, normalized to its peak value, for increasing values of $\Delta z_{\rm img}$, showing the rapid deterioration of the X-shape of correlation.  The decrease of the peak height (normalized to the 
$\Delta z_{\rm img}=0$ value) is shown in (b). The plot (c) shows the cross-section  of the central peak along the temporal delay $\Delta t$, evidencing a broadening of the temporal correlation for increasing errors in the imaging plane 
}
\label{fig_dzimg2}
\end{figure*}
Figure \ref{fig_dzimg2} plots the spatio-temporal profile of the detected PDC correlation $|\psiout(\Delta x,0,\Delta t)|^2$ (row a and b) and of its temporal cross-section at $\Delta x=0$ (raw c) for increasing values of the error $\Delta z_{\rm img}$, evaluated from the PWPA result (\ref{Icoh1}). 
The density plots in row a) shows that a large broadening occurs along the spatial direction. This  is not unexpected:  with to respect an error in the imaging plane,  the reconstructed spatial correlation as a function of $\Delta x$ behaves as the spatial  resolution of an ordinary optical image with respect to the depth of focus of the imaging system, i.e. it roughly broadens as 
$\Delta x_{\rm fwhm}\sqrt{1+\Delta z_{\rm img}^2/z_{\rm DOF}^2}$.\\ 
Less expected is perhaps the broadening of the correlation along the temporal direction, evidenced by raw c) in Fig.\ref{fig_dzimg2}. This temporal broadening is a clear consequence of the non-factorability of the twin-beam correlation in space and time. In turns, this is a consequence of the non-factorable character of phase matching: the phase matching condition [see Eq. \eqref{pm2}] 
\beq
\Deltapdc(q,\Omega)\approx\Omega^2/\Omega_D^2-q^2/q_D^2=0
\leftrightarrow k_1''\Omega^2=q^2/k_1
\label{pm0}
\eeq
can indeed be read as a compensation of the temporal dispersion experienced by twin photons inside the PDC crystal due to diffraction. 
Only if the entrance face of the second crystal is placed exactly in the image plane of the first crystal, this compensation  occurs
and the biphoton correlation is a nearly transform limited coherent sum of the phase-matched spectral modes.\\
On the contrary, if the second crystal is misplaced with respect to the image plane, free propagation
deteriorates this exact compensation of dispersion and diffraction. 
By making the simple assumption that only phase matched modes contribute to the coherent SFG
component (\ref{Icoh1}), we can make the substitution $q^2\rightarrow\Omega^2 k_1 k_1''$ 
in the propagation term at the r.h.s. of \eqref{phase_dz_img}.  
It transforms then into the $\Omega-$dependent phase factor $e^{-i\frac{n_1 k_1''\Omega^2}{1-\Omega^2/\omega_1^2}\Delta z_{\rm img}}$
which describes a quadratic dispersion-like chirp of the twin beams.

\subsection{Effect of misalignments of the two crystals} 
\begin{figure}[ht]
\centering
{\scalebox{.55}{\includegraphics*{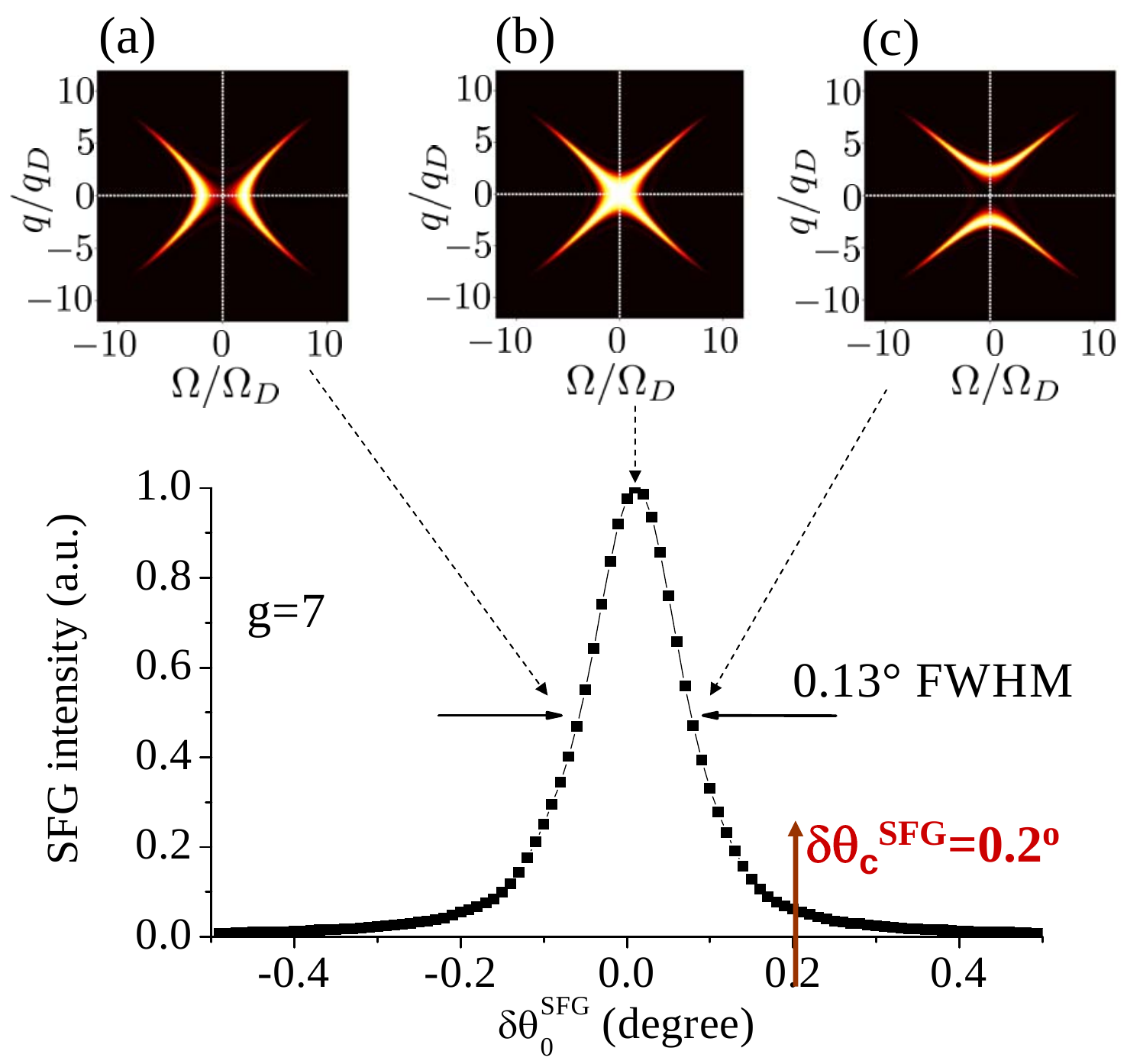}}}
\caption{(color online) Bottom: peak of the coherent SFG intensity $I_{\rm SFG}^{\rm (coh)}[0,0,0]$ as a function of the angular tilt $\Delta \theta_0^{\rm SFG}$ between the two crystals (normalized to the value it has for $\delta \theta_0^{\rm SFG}=0$). Top: plot of $|\Sicoh(\q,\Omega)|^2$ for three different orientations
of the SFG crystal: (a) $\delta \theta_0^{\rm SFG}=-0.06^\circ$, $\Delta_0^{\rm SFG}=-4.24$, (b) $\delta \theta_0^{\rm SFG}=0^\circ$,
$\Delta_0^{\rm SFG}l_c'=0$, and (c) $\delta \theta_0^{\rm SFG}l_c'=0.07^\circ$, $\Delta_0^{\rm SFG}l_c'=5.8$.}
\label{tilt_SFG}
\end{figure}

The coherent component of the SFG light is also strongly sensitive to misalignments
of the SFG crystal with respect to the PDC crystal orientation.  
We investigated numerically the effect of a small tilt 
$\delta \theta_0^{\rm SFG}\neq 0$ of the SFG crystal with respect to the PDC crystal orientation, the latter 
satisfying perfect phase-matching at degeneracy, i.e.  $\Delta_0^{\rm PDC}=0$.
Figure \ref{tilt_SFG} shows how the peak of the coherent contribution , i.e
[see Eqs.(\ref{IcohA})-(\ref{psi_out})]:
\beqa
\label{Icoh_peak}
&&\Icoh[0,0,0]
  =
  \left|
   \int\frac{d\vka}{(2\pi)^3}
   \UV(\vka)
   \Sicoh(\vka)
   \right|^2, 
\eeqa
rapidly goes to zero as the tilt angle exceeds a few tenths of degrees.
As expected, $I_{\rm SFG}^{\rm (coh)}$ takes its maximum value when the two crystals are perfectly aligned, as $\UV$ and $\Sicoh$,
are peaked around  the same phase-matching curves and their overlap integral is maximized.  

We can give a quantitative estimate of the angular tolerance in the alignment of the two crystals 
by considering explicitly the case  when the second crystal is slightly tilted with respect 
to the first crystal, the latter being tuned for collinear phase matching. 
Denoting with $\theta^{\rm PDC}$ $(\theta^{\rm SFG})$ the orientation angle of the PDC (SFG) crystals with the pump axis
(i.e. the $z$-axis), 
and assuming the tilt angle between the two crystals $\delta \theta^{\rm SFG}= \theta^{\rm SFG}-\theta^{PDC}$ is small, 
the following approximate relation between the pump mode wavenumbers in the two crystals holds 
\beq 
k_0^{\rm SFG} \approx  k_0 
-k_0\rho_0\delta \theta^{\rm SFG} 
\label{k0sfg}
\eeq 
where $\rho_0= -\partial k_0 /\partial _{q_x} |_{\omega=\omega_0, q=0}$  is the positive defined walk-off angle of the 
extraordinary wave at frequency $\omega_0$. 
Since by assumption $\Delta_0^{\rm PDC}\equiv 2k_1-k_0=0$,
according to Eq.(\ref{k0sfg}) a negative tilt $\delta \theta^{\rm SFG}<0$  leads
to a positive  collinear phase-mismatch $\Delta_0^{\rm SFG}\equiv2k_1-k_0^{\rm SFG}=-k_0\rho_0\delta \theta^{\rm SFG}>0$ 
in the second crystal. At degeneracy the phase-matched spatial modes in the second crystal lye therefore on 
a circumference of radius 
\beq
q_R=q_D \sqrt{\Delta_0^{\rm SFG} l_c'}\approx q_D \sqrt{k_0\rho_0 l_c' |\delta\theta^{\rm SFG}|}.
\label{q_radius}
\eeq 
Close to degeneracy, the region of overlap of $\UV$ and $\Sicoh$ can be expected to reduce drastically when $q_R$ 
coincides with the first  node of the PDC probability amplitude $\UV(\q,\Omega=0)$ along the spatial frequency axis,
which is found to be well approximate by $\bar{q}_D=\sqrt{2}(\pi^2+g^2)^{1/4}q_D$ [see Eq.\eqref{PDC_band_g}]. 
Following this criterium, the overlap integral (\ref{Icoh_peak}) will be reduced by a large amount when 
the tilt angle exceeds the critical value 
\beq
\label{angular_tol}
\delta\theta_c^{\rm SFG}
=\frac{2\sqrt{\pi^2+g^2}}{\rho_0 k_0 l_c'}
\eeq
Taking e.g. $g=7$ it gives us a tolerance $\delta\theta_c^{\rm SFG}=0.2^\circ$ in the alignment of the second BBO crystal
with respect to the first. This value is in agreement with the numerical evaluation of $I_{\rm SFG}^{\rm (coh)}[0,0,0]$
as a function of $\delta\theta_0^{\rm SFG}$
shown in Fig.\ref{tilt_SFG}.


\section{Conclusions}
Our treatment shows that the SFG process represents a powerful tool for exploring the biphotonic correlation in the full spatio-temporal domain. Through a careful manipulation of both the spatial and the temporal degrees of freedom of the PDC field emitted by the first 
$\chi^{(2)}$ crystal, the proposed optical setup allows to retrieve
the strongly localized X-shaped PDC correlation in the space time domain,
predicted in \cite{gatti2009,caspani2010,brambilla2010}. 
The analytical result obtained within the PWPA shows that the coherent component of the up-converted SFG field contains
the desired information on the PDC correlation function. 
In particular, this coherent component can be expressed in the form of a convolution
between the PDC biphotonic function in direct space, the quantity under investigation, with the corresponding probability
amplitude describing the up-conversion process in the SFG crystal, the convolution being evaluated
at the applied temporal delay and spatial shift between two conjugate PDC components.
It is shown that the measured quantity retains the main features of the biphotonic correlation, namely its
nonfactorable X-shaped geometry and its strong localization in space and time. The latter, which can in principle reduce to
a few pump optical cycles \cite{gatti2009}, is determined by the acceptance bandwidth of the up-conversion process.
Finally, a fully 3D+1 numerical modeling of the optical setup that takes into account the pump pulse finite size 
has been implemented in order to provide a more realistic simulation of the experiment being implemented in Como. 
This allowed us to show that, even in the high gain regime of PDC, the visibility of the correlation
measurement con be close to 100\% when the up-converted light is collected in the far-field of the SFG crystal, and
for evaluating the tolerance of the phase-sensitive correlation measurement against imperfection of the imaging system.
\acknowledgments
We are grateful to Paolo di Trapani for  precious suggestions and discussions. 
This work was realized in the framework of the Fet Open project of EC 221906 HIDEAS. 

\appendix
\setcounter{equation}{0}
\section{Derivation of Eqs.(\ref{inout1})-(\ref{transferH})}
\label{appendixA}
\begin{figure}
{\scalebox{.7}{\includegraphics*{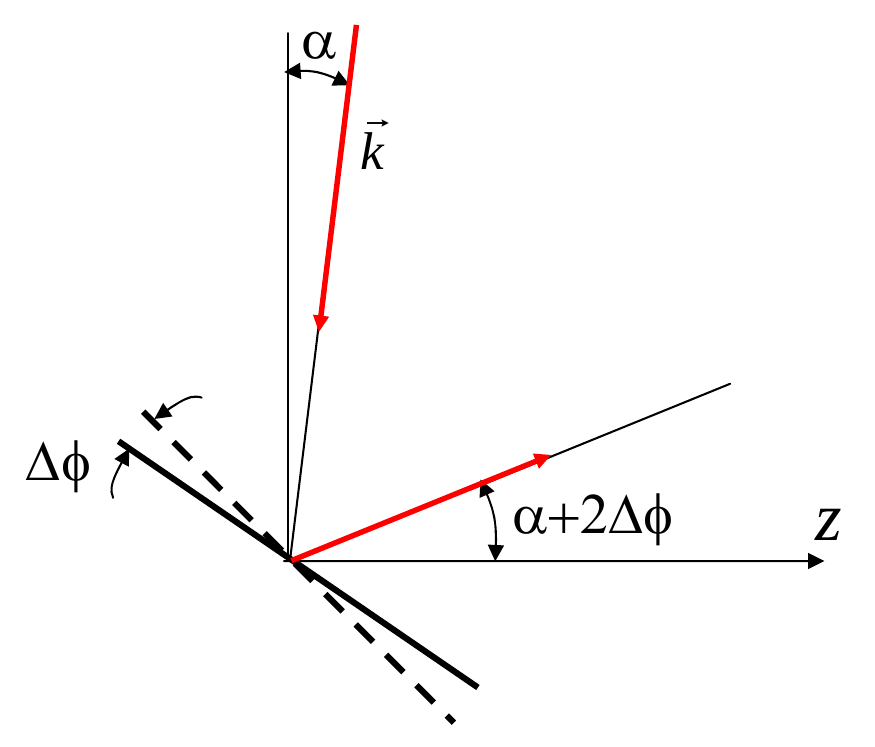}}}
\caption{(color online) A rotation of $\Delta \phi$ of the mirror with respect to its original position (dashed line) generates 
a tilt by $2\Delta \phi$ of the $k$-vector with respect to the $z$-axis direction after reflection.}
\label{fig_M1}
\end{figure}

In this appendix we describe the effect of a rotating mirror set in the 2f-plane of a 4-f telescopic system.
Let us consider a rotation by $\Delta \phi$ of the mirror across the $y$-axis (orthogonal to the figure plane)
with respect to its normal position set at $45^\circ$ with respect to the incident beam (dashed line in Fig.\ref{fig_M1}).  
Because of reflection, the $k$-vector of an incident mode $(\q,\Omega)$ with free space wavenumber 
$k_f(\Omega)=(\omega_1+\Omega)/c$ 
and direction angle $\alpha=\arcsin(q_x/k_f)$ in the $(x,z)$-plane is tilted by an additional angle $2\Delta\phi$
with respect to the $\Delta\phi=0$ direction.
Accordingly, the $k$-vector transverse and longitudinal component in the plane orthogonal to the
rotation axis, $q_x=k_f\sin\alpha$ and $k_z=k_f\cos\alpha$,
transforms according to the orthogonal transformation
\bsub
\label{k_rot}
\beqa
q_x'&=&q_x \cos2\Delta\phi  +k_{fz}\sin 2\Delta \phi \\
k_{fz}'&=&k_{fz} \cos2\Delta\phi  -q_x\sin 2\Delta \phi
\eeqa
\esub
For small angles, i.e. for $q_x/k_f=\sin\alpha\ll 1$ and $\Delta \phi\ll 1$, relations (\ref{k_rot})
reduces to 
\beq
q_x'=q_x+2 k_f \Delta\phi,\;\;k_{fz}'=k_{fz}
\eeq
where all powers of $q_x/k_z$ and $\Delta \phi$ have been neglected except the linear terms.
Under this approximation the relation between the reflected and the incident field, $e_{1r}$ and $e_{1i}$, can be written as 
\beq
e_{1r}(q_x,q_y,\Omega)=e_{1i}(q_x-2 k_f(\Omega) \Delta \phi,q_y,\Omega)
\label{M1inout}
\eeq
The field Fourier transformation performed by the second lens of the $4-f$ system
(see Fig.\ref{fig1}) can be written as
\beqa
c_{1}(\x,\Omega)&=&-\frac{ik_f(\Omega)}{2\pi f}
\int d\xp e^{-i\frac{k_f(\Omega)}{f} \x \cdot\xp} e_{1r}(\x,\Omega)\nn\\
&=&-\frac{ik_f(\Omega)}{f} 
e_{1r}\left(\q=\frac{k_f(\Omega)}{f}\x,\Omega\right)\;,
\label{lens1}
\eeqa
while that performed by the first lens can be written as
\beq
e_{1i}(\q,\Omega)=-\frac{if}{k_f} b_1\left( \x=-\frac{f}{k_f}\q,\Omega\right)\;.
\label{lens2}
\eeq
Combining Eqs.(\ref{M1inout})-(\ref{lens2}) together, we readily obtain the following transformation between the input and the output planes 
of the telescopic system
\beq
c_1(\x,\Omega)=-b_1(-\x+\Delta x,\Omega)
\label{M1_xshift}
\eeq
where $\Delta \x=(2f\Delta \phi,0)$ denotes the transverse displacement of the field at the SFG crystal input plane 
produced by the mirror rotation. It is worth noticing that, under this approximation, the transverse
shift is the same at all temporal frequencies.

In the Fourier domain, Eq.(\ref{M1_xshift}) reduces to the linear phase-shift transformation 
\beq
c_1(\q,\Omega)=-e^{i\q\cdot\Delta x}b_1(-\q,\Omega)\;,
\eeq
which coincides with input-output relation (\ref{M1_rot}) affecting the $q_x>0$ modes of beam +,
except for the $\q\rightarrow-\q$ reflection and the $-1$ factor introduced by the two lenses 
which has been omitted in the main treatment.

\end{document}